\documentclass[a4paper]{article}

\usepackage[utf8]{inputenc}

\usepackage{subfigure}
\usepackage{epic,eepic}
\usepackage{graphicx,color}

\usepackage{float}
\usepackage{ifthen}
\usepackage{xspace}

\usepackage{amssymb}
\usepackage{amsmath}
\usepackage{amsfonts} 
\usepackage{stmaryrd} 
\usepackage{latexsym} 

\newtheorem{notation}{Notation}

\newtheorem{definition}{Definition}

\newcommand\set[1]{\ensuremath{\{ #1 \} }}
\renewcommand\int[1]{\ensuremath{\llbracket #1 \rrbracket}}

\newcommand\comment[1]{}

\renewcommand{\textbf}[1]{\begingroup\bfseries\mathversion{bold}#1\endgroup}

\begin{document}

\title{The Link Stream of Contacts in a Whole Hospital\thanks{This study was supported by the European Commission under the Life Science Health Priority of the 6th Framework Program (MOSAR network contract LSHP-CT-2007-037941).
This work was performed within the framework of the LABEX MILYON (ANR-10-LABX-0070) of Universit{\'e} de Lyon, within the program "Investissements d'Avenir" (ANR-11-IDEX-0007) operated by the French National Research Agency (ANR).
This work was performed within the framework of the LABEX IBEID (ANR-10-LABX-62).
The second author gratefully acknowledges the support from a grant from R{\'e}gion Rh{\^o}ne-Alpes and from the delegation program of CNRS.}
\thanks{This is a complete and augmented version of the extended abstract appeared in CompleNet 2014~\cite{MCF14}.}
}

\author{
Lucie Martinet
\thanks{Univ Lyon, ENS de Lyon, UCB Lyon 1, Inria, CNRS, LIP UMR 5668, 15 parvis Ren{\'e} Descartes, F-69342 Lyon, FRANCE -- \texttt{lucie.martinet@ens-lyon.org}}
\and
Christophe Crespelle
\thanks{Univ Lyon, ENS de Lyon, UCB Lyon 1, Inria, CNRS, LIP UMR 5668, 15 parvis Ren{\'e} Descartes, F-69342 Lyon, and Institute of Mathematics, Vietnam Academy of Science and Technology, 18 Hoang Quoc Viet, Hanoi, VIETNAM FRANCE -- \texttt{christophe.crespelle@inria.fr}}
\and
Eric Fleury
\thanks{Univ Lyon, ENS de Lyon, UCB Lyon 1, Inria, CNRS, LIP UMR 5668, 15 parvis Ren{\'e} Descartes, F-69342 Lyon, FRANCE -- \texttt{eric.fleury@inria.fr}}
\and
Pierre-Yves Bo{\"e}lle
\thanks{Sorbonne Universit{\'e}s, UPMC Univ Paris 06, INSERM, Institut Pierre Louis d'{\'E}pidemiologie et de Sant{\'e} publique, UMR S 1136 -- \texttt{pierre-yves.boelle@upmc.fr}}
\and
Didier Guillemot
\thanks{Institut Pasteur, Unit{\'e} de Pharmaco-{\'E}pid{\'e}miologie et Maladies Infectieuses, INSERM, U1181, and Univ. Versailles Saint Quentin, UFR des Sciences de la Sant{\'e} Simone-Veil, EA 4499, and AP-HP, H{\^o}pital Raymond-Poincar{\'e}, Unit{\'e} Fonctionnelle de Sant{\'e} Publique -- \texttt{didier.guillemot@uvsq.fr}}
}

\date{}

\maketitle

\begin{abstract}
We analyse a huge and very precise trace of contact data collected by a network of sensors during 6 months on the entire population of a rehabilitation hospital. We investigate both the topological structure of the average daily link stream of contacts in the hospital and the temporal structure of the evolution of these contacts hour by hour. Our main results are to unveil striking properties of these two structures in the considered hospital, and to present a methodology that can be used for analysing any link stream where nodes are classified into groups.
\end{abstract}


\maketitle




\section*{Introduction}

The prevalence of AMRB (AntiMicrobial Resistant Bacteria) has been rising worldwide during the past decades and the resistance rates for major nosocomial pathogens have increased up to alarming levels, implying adverse outcomes for affected patients, such as delays or failures of therapies, prolonged hospitalization stay and increased mortality.
Upon colonization by an AMR bacteria, a patient becomes an \emph{occult carrier}. He is then a potential colonization source for other patients and may also disseminate AMRB into the community while transferred to other facilities. In this context, rehabilitation centres are considered to be a large reservoir of AMRB, offering a great potential for development and dissemination into the community.

The MOSAR project aims at examining the factors determining the dynamics of AMRB spread within healthcare facilities~\cite{OSO+15,OOT+15}. These factors are numerous and complex, but it is widely believed that one support for transmissions of AMRB is close proximity interactions (which we simply call \emph{contacts} throughout this article)~\cite{Hal06,EKW+06,MHJ+08,MAR+08,BFM+09,GMB+09,TOB+09,PTP+10}. Then, to further reduce transmission, in addition to classical prevention measures~\cite{MJ+03} (such as admission controls, isolation of carriers and hand hygiene), controlling the flux of interactions within the hospital is considered as the next step \cite{DS+04,WS+05,NBW05}. Indeed, contacts and their dynamics strongly influence how transmission occurs~\cite{PV01,Moo02,RK03,Eam08,SFS09,SJ10,SV+11b}. Yet, contacts are difficult to measure efficiently in practice~\cite{BSA+06,RER+12}, and they may even be harder to change. Recently, however, advances in communication technologies have made it possible to record close proximity interactions with unprecedented detail, allowing an in depth view of the structure of contacts in real-life settings~\cite{HCS+15,EP06,CV+10,LL+12} including in environments critical for spreading of diseases~\cite{SKL+10,SV+11,IR+11,HN+12,VBC+13}. If such contacts actually support transmission, it may open the way to further improvement in hospital hygiene policy. 

In this article, we analyse the contact trace collected on the entire population of a rehabilitation hospital during 6 months between June and November 2009, within the MOSAR project. We focus on a period of 59 days (a bit more than 8 weeks) of the measurement, from July 6th to September 2nd involving 492 individuals, 253 patients and 239 staffs. We describe the methodology we used to uncover the key characteristics of this link stream of contacts and the main results we obtained.

\begin{figure}
\centering
\includegraphics[width=\linewidth]{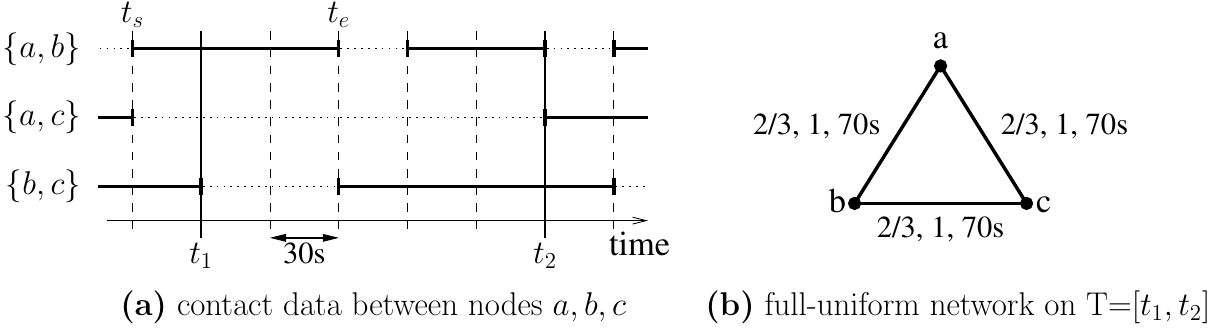}
\caption{Sample contact data and full-uniform network. (a) sample contact data between three nodes $a,b,c$, (b) the full-uniform network formed from this contact data on period $T=[t_1,t_2]$. The schema shows 7 contacts between nodes $a,b,c$, e.g. the first contact between $a$ and $b$ lasts 90s from time $t_s$ to time $t_e$. On the period $T$ of study, there are 2 adjacency pairs, namely $\set{a,b}$ and $\set{b,c}$ (nodes $a$ and $c$ are never in contact between $t_1$ and $t_2$), 3 contacts (2 between nodes $a,b$ and 1 between nodes $b,c$) and the cumulated length of these contacts on period $T$ is 210s (60s for each of the 2 contacts between nodes $a,b$ and 90s for the contact between nodes $b,c$). Then, in the full-uniform network formed on period $T$, the values associated with each of the 3 couples of nodes are 2/3 for the number of adjacency pairs, 3/3=1 for the number of contacts and 210/3=70s for the cumulated length of contacts.
\label{fig:ex-contact}}
\end{figure}

\subsection*{Our contribution}

We analyse separately the graph structure of the average daily link stream of contacts (without taking into account its evolution over time) and the temporal evolution of contacts in the hospital hour by hour.
For the first goal, we point out significant differences in the contact profiles of services
, as well as in contact patterns of patients and staffs
, and we reveal a very special structure of interconnections between the services of the hospital
and between the socio-professional categories
.
Finally, 
we show that the temporal evolution of the contacts in the hospital presents a clear circadian and weekly pattern, and we unfold the very different behaviours of patients and staffs in this temporal pattern.

\begin{figure}
\centering
\includegraphics[width=0.55\linewidth]{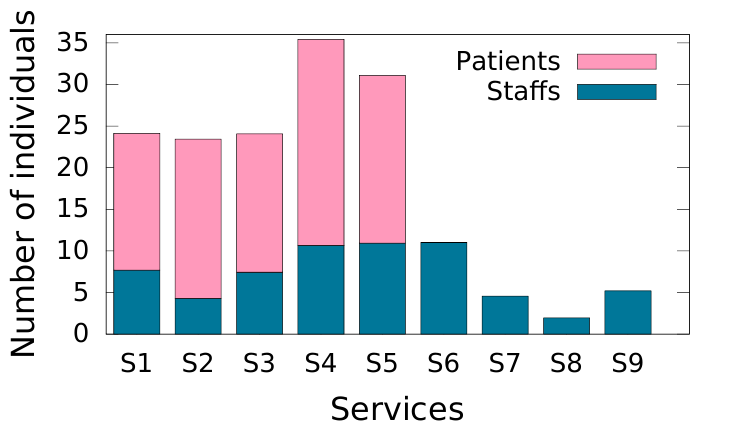}
\caption{Mean number of individuals per day in each service. Patients are plotted in light pink and staffs in dark turquoise.}
\label{repart-serv}
\end{figure}

\begin{table}
\normalsize
\begin{tabular}{l|l|}
\hline
S1 (Sorrel) & nutritionnal readaptation \\
S2 (Sorrel) & neuro-orthopedic reeducation \\ 
S3 (Sorrel) &  geriatric readaptation \\
S4 (Ménard) & chronic vegetative state \\
S5 (Ménard) & neurologic readaptation \\ 
\hline
\end{tabular}
\begin{tabular}{l|l|}
\hline
S6 & night service \\
S7 & physiotherapy \\
S8 & ergotherapy \\
S9 & other reeducation staffs\\
 & \\
\hline
\end{tabular}
\caption{Functions of the nine services of the hospital.}
\label{tab:serv}
\end{table}

\subsection*{Related works}
There have been several recent works using sensor devices in order to unfold contact patterns among individuals (both graph structure and temporal structure) in environments involving patients or children, which present critical risks for spreading of diseases. The measurement analysed in \cite{SV+11} was made on an entire primary school during three days. The experiments described in \cite{IR+11,HN+12} were both conducted during one week in some paediatric ward and the one of~\cite{VBC+13} took place in a geriatric ward, a kind of service we also have in our study, during three days. Finally, for sake of completeness, let us mention that a similar experiment was recently conducted on part of the population of an office building during two weeks~\cite{GVF+15}.
Compared to those works, our analyses present two important advantages. Firstly, the measurement we use was made on a much longer period of time (6 months), which allows to observe weekly pattern and to assess the generality of the conclusions we derive on shorter period of times (like one day or one week). Secondly, our measurement is not limited to a specific part of the hospital, it involves all patients and all staffs\footnote{More than 99\% of individuals accepted to participate and register to the experiment.} of all services of the hospital, which is a key point to have an accurate view of the actual possibility of spreading into the whole hospital, and even inside a given service. Indeed, these possibilities also depend on the contacts occurring outside the service under study.

\begin{figure}
\centering
\includegraphics{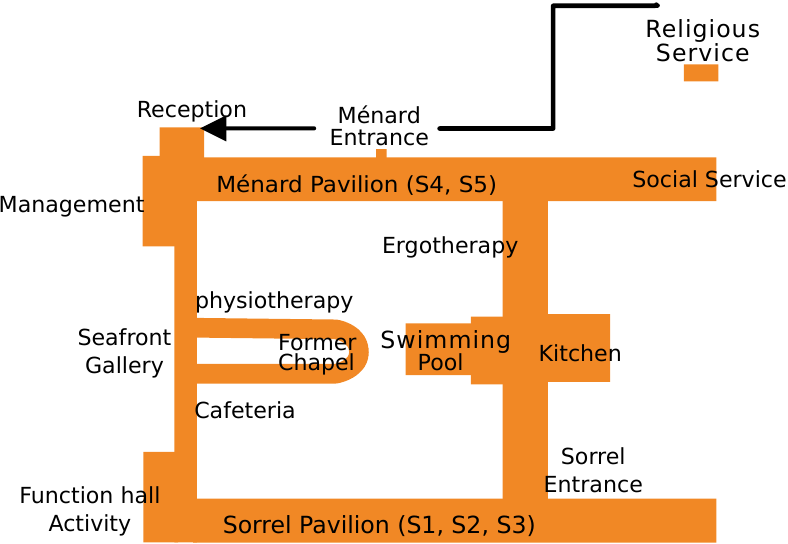}
\caption{Plan of the hospital of Berck-sur-Mer. With locations of healthcare services.}
\label{fig:plan}
\end{figure}

\section{Preliminaries}\label{sec:prel}

The contact data we analyse here was recorded using wireless sensor devices carried by all the participants. Each device $a$ sends one beacon signal containing the ID of $a$ every 30s and constantly listens for the signals of other devices\footnote{The sending time of the different devices are not synchronised but their internal clocks are.}. As a consequence, the signal sent by device $a$ is received by all the other devices $b$ that are close enough from $a$ (typically 1 to 1.5 meters). When this occurs, $b$ records the ID of $a$ together with the timestamps of the reception of the signal. On the technical side, it is worth to note that this short transmission range is obtained by sending low-power signals, which also implies that only signals between sensors that are not separated by a physical obstacle, such as a wall, a door or even a human body, are recorded.

Afterwards, time is sliced in slots of 30s and we keep, for each slot, the list of pairs $\set{a,b}$ of sensors such that one ($a$ or $b$) recorded the signal of the other. These pairs are undirected as we do not keep track of whether $a$ received the signal of $b$ or $b$ received the signal of $a$ or both: all these three situations give rise to one single occurrence of the (undirected) pair $\set{a,b}$ in the considered 30s time slot.

Finally, if a pair $\set{a,b}$ occurs in several consecutive time slots of 30s, we group all its consecutive occurrences into one single interval of contact. Consequently, the contact data analysed in this article consists of a set of quadruplets $(a,b,t_s,t_e)$, which is called a \emph{link stream} and is denoted $L$ in the following (see~\cite{LVM17} for definitions and terminology). For each quadruplet $(a,b,t_s,t_e)\in L$, $a$ and $b$ are two nodes, $t_s$ is the starting time of one 30s time slot and $t_e$ is the ending time of one 30s time slot, with the additional condition $t_s<t_e$ (see example in Figure~\ref{fig:ex-contact}). The meaning of quadruplet $(a,b,t_s,t_e)$ is that nodes $a$ and $b$ are in contact during all the time slots between $t_s$ and $t_e$ and that they are not in contact in the 30s time slot immediately preceding $t_s$ as well as in the 30s time slot immediately following $t_e$. This is why in the present context, each quadruplet in $L$ is called a \emph{contact} and has a non-null \emph{length}, namely $t_e-t_s$, which is a multiple\footnote{Note that, because of the way we slice the time in slots of 30s, the condition "$t_1-t_2$ is a multiple of 30s" holds for any $t_1$ and for any $t_2$ being the bound of some interval, even if they do not bound the same interval of contact.} of 30s. The author can refer to~\cite{OSO+15} for a more detailed description of how contact data was gathered.

\begin{figure}		
\centering
\includegraphics[width=\linewidth]{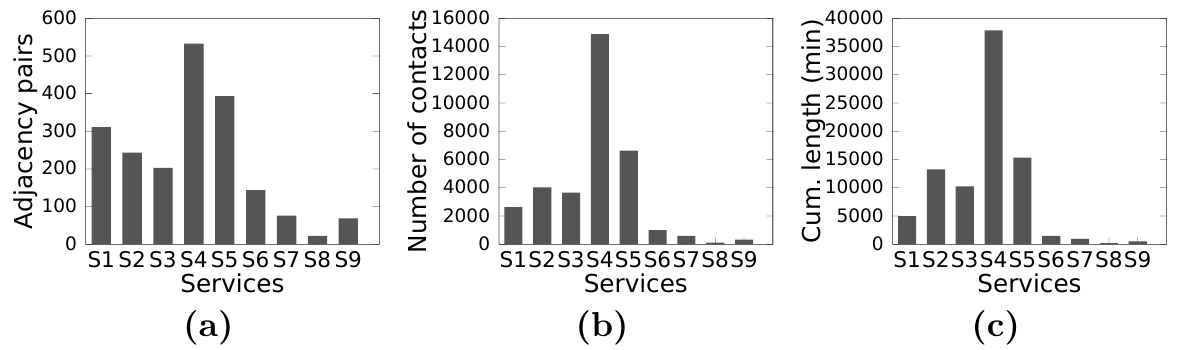}
\caption{Mean activity per day for each service. (a) number of adjacency pairs, (b) number of contacts, (c) cumulated length of contacts.}
\label{TotalContacts}
\end{figure}

\begin{figure}		
\centering
\includegraphics[width=\linewidth]{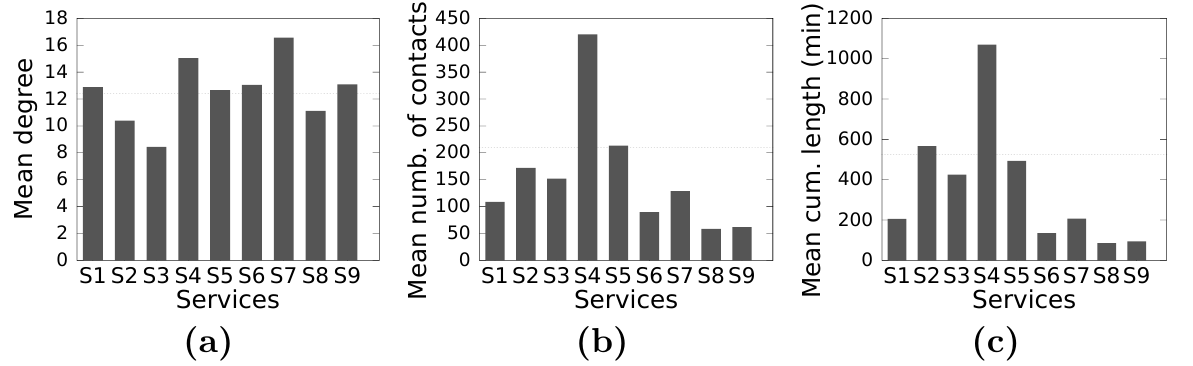}
\caption{Mean activity per individual and per day for each service. (a) number of adjacency pairs, (b) number of contacts, (c) cumulated length of contacts. The doted lines depict the mean values per individual in the whole hospital.}
\label{TotalContIndiv}
\end{figure}

\begin{notation}[Adjacency pair and length of a contact]
For a contact $C=(a,b,t_s,t_e)$, we denote $pair(C)=\set{a,b}$ the pair of nodes involved in contact $C$, which we call the \emph{adjacency pair} of $C$, and $length(C)=t_e-t_s$ the length of contact $C$.
\end{notation}

\begin{notation}[Set of adjacency pairs and set of nodes of a linkstream]
For a link stream $L$, we denote $E(L)=\set{pair(C) |\ C\in L}$ the set of adjacency pairs of $L$ and $V(L)=\bigcup_{C\in L} pair(C)$ the set of nodes involved in $L$.
\end{notation}

Throughout the article, we often analyse link streams restricted to a specified time period (typically one day or one hour), which are formally defined as follows.

\begin{definition}[Link stream $L(T)$ restricted to time period $T$]\label{def:timeres}
\begin{sloppypar}
The link stream $L(T)$ obtained from $L$ by restriction to the time period $T=[t_1,t_2]$ is defined as $L(T)=\set{(a,b,t_s,t_e)\ |\ t_s\neq t_e \text{ and } \exists (a,b,t'_s,t'_e)\in L, [t_s,t_e]=[t'_s,t'_e]\cap [t_1,t_2]}$.
\end{sloppypar}
\end{definition}

\begin{figure}
\centering
\includegraphics[width=\linewidth]{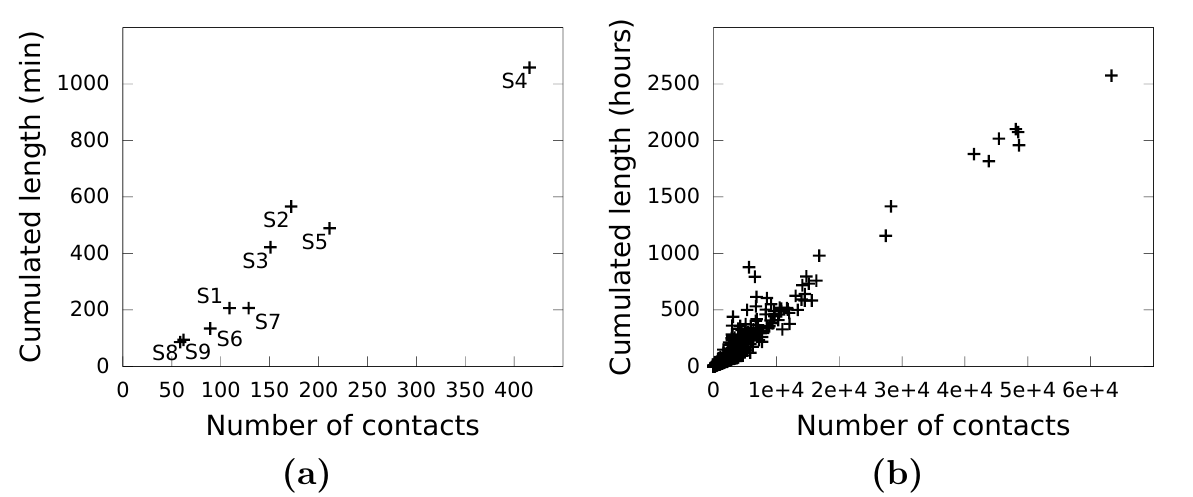}
\caption{Scatter plots of the number of contacts vs the cumulated length of contacts. (a) mean values per individual and per day for each service, (b) values for each individual in the hospital on the whole period of experiment.}
\label{EquivContLength}
\end{figure}

We decline our analyses using three parameters.

\begin{definition}
For a link stream $L$, we define the three following parameters:
\begin{enumerate}
\item number of adjacency pairs, $\#pairs(L)=|pairs(L)|$,
\item number of contacts, $\#cont(L)=|L|$ and
\item cumulated length of contacts, $cumul\_length(L)=\sum_{C\in L} length(C)$.
\end{enumerate}
\end{definition}

In the following, we usually use these parameters for link streams obtained by restriction of a larger link stream to a time period (see example in Figure~\ref{fig:ex-contact}). Moreover, we often use the above notions restricted to the contacts of a designated group of individuals or to the contacts between one single pair of nodes. Moreover, in the rest of the article, we use the notions of \emph{semi-contact} and \emph{adjacency semi-pair} instead of contact and adjacency pair. One contact (resp. one adjacency pair) between $a$ and $b$ gives rise to two \emph{semi-contacts} (resp. two \emph{adjacency semi-pairs}): one attached to $a$ and one attached to $b$. For sake of vocabulary simplicity, in the following, we use terms contact and pair instead of semi-contact and semi-pair, but all statistics are actually made using semi-contacts and semi-pairs. The reason for this is that it gives a straightforward meaning to mean statistics per individual.

In order to analyse the graph structure of contacts within the hospital, we use the aggregated view of a link stream defined below. 

\begin{definition}[Aggregated network of a link stream]\label{def:aggreg}
The aggregated network $G$ of a link stream $L$ is the graph $G=(V(L),E(L))$ made of the adjacency pairs of $L$. Moreover, in the aggregated network $G$, each adjacency pair $\set{u,v}$ is given two weights denoted $\#cont_{\set{u,v}}$ and $cumul\_length_{\set{u,v}}$ and defined as $\#cont_{\set{u,v}}=|\set{C\in L\ |\ pair(C)=\set{u,v}}|$ and $cumul\_length_{\set{u,v}}=\sum_{C\in L \text{ and } pair(C)=\set{u,v}} length(C)$.
\end{definition}

\begin{figure}	
\centering
\includegraphics[width=\linewidth]{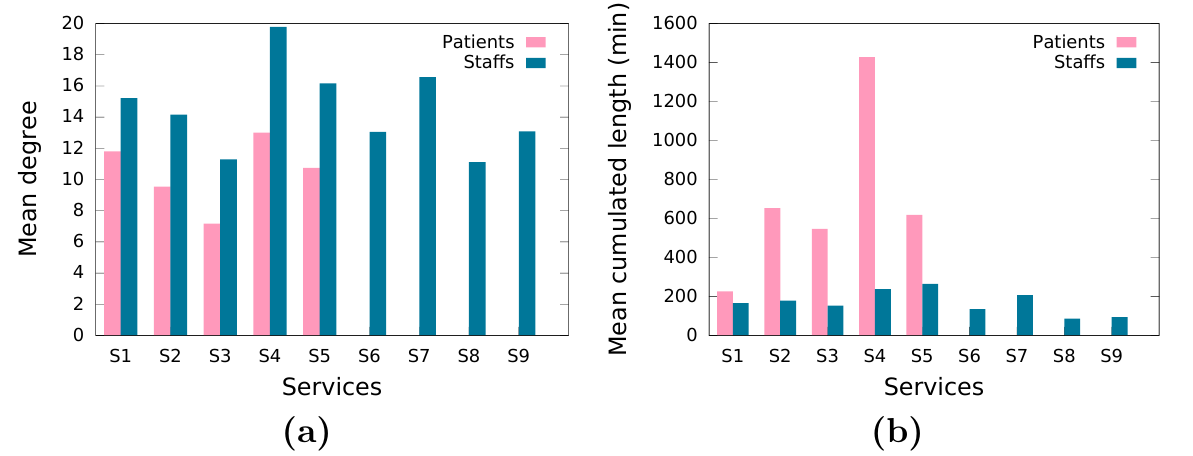}
\caption{Mean activity per individual and per day for each service. (a) number of adjacency pairs, (b) cumulated length of contacts. Patients are in light pink and staffs in dark turquoise.}
\label{TotalContactsPAPE}
\end{figure}

Along this article, for sake of comparison, we make extensive use of a uniformised version of the aggregated network of the hospital, which we call the \emph{full-uniform network} and which is defined as follows (see example in Figure~\ref{fig:ex-contact}).

\begin{definition}[Full-uniform network]\label{def:fulluniform}
The full-uniform network associated to a link stream $L$ is the complete graph on vertex set $V(L)$ where each pair of nodes $u,v\in V(L)$ receives three quantities, which have the same value for all pairs of nodes of $V(L)$:
\begin{enumerate}
\item a fractional number $\#adj(u,v)$ of adjacency pairs, between $0$ and $1$, equal to the density of adjacency pairs between nodes of $V(L)$, i.e. $\#adj(u,v)=2\, \#pairs(L)/(|V(L)|$ $(|V(L)|-1))$,
\item a number of contacts $\#cont(u,v)$ equal to the mean number of contacts per pair of nodes, i.e. $\#cont(u,v)=2\, \#cont(L)/(|V(L)|(|V(L)|-1))$,
\item a cumulated length of contact $cumul\_length(u,v)$ equal to the mean cumulated length per pair of nodes, i.e. $cumul\_length(u,v)=2\, cumul\_length(L)$ $/(|V(L)|$ $(|V(L)|-1))$.
\end{enumerate}
\end{definition}

\subsection*{General organisation of the hospital}

Over the period of study, the mean number of people present in the hospital during one day is about 103 patients and 64 staffs.
The patients and staffs are divided into 9 services. The average daily size of each service is given in Figure~\ref{repart-serv} and their functions are given in Table~\ref{tab:serv}. Only the first five of them (S1 to S5) are hospital wards and contain both patients and staffs, the other four (S6 to S9) contain only staffs. Each of the 5 services involving patients occupy one floor in one of the two wings (called Ménard and Sorrel) of the building (see map in Figure~\ref{fig:plan}). The Sorrel wing hosts three services: nutritional readaptation (S1, 1st floor), neuro-orthopedic reeducation (S2, 2nd floor), geriatric readaptation (S3, 3rd floor). The Ménard wing hosts the chronic vegetative state service (S4, 2nd floor) and the neurologic readaptation service (S5, 3rd floor).
The four services containing only staffs are the night service (S6), regrouping people replacing staffs from services S1 to S5 during nights, and the reeducation services (S7 to S9): physiotherapeutic service (S7), ergotherapeutic service (S8) and other reeducation staff service (S9). S7 and S8 are located in two distinct places between the two wings of the buildings, but S6 and S9 do not have a determined location in the hospital. It must be clear that the division of the hospital into services is not meaningful only from an administrative and management point of view but also clearly impacts the structure of the contacts: in average in one day, 66\% of the adjacency pairs of the hospital occur inside services, 88\% for the number of contacts and 92\% for the cumulated length of contacts, while these values are only 25\% in the full-uniform network.

\begin{table}

\normalsize
\centering
\begin{tabular}{c}
\begin{tabular}{c|c|c|c}
 &	PA-PA	&PA-ST&	ST-ST\\
 \hline
Pairs&	0.24&	0.56&	0.20\\
Length&	0.80&	0.12&	0.08\\
\end{tabular}
\\[0.6cm]
(a) Global distribution
\end{tabular}


\smallskip

\null\hfill
\begin{tabular}{c}
\begin{tabular}{c|c|c}
PA vs &PA	&ST	\\
\hline
Pairs	&0.46	&0.54 	\\
\hline
Length	&0.93	&0.07	\\
\end{tabular}
\\[0.6cm]
(b) Patients centred
\end{tabular}
\hfill
\begin{tabular}{c}
\begin{tabular}{c|c|c}
ST vs	&PA	&ST\\
\hline
Pairs	&0.58	&0.42	\\
\hline
Length	&0.42	&0.58\\
\end{tabular}
\\[0.6cm]
(c) Staffs centred
\end{tabular}
\hfill
\null
\label{GenStatPAPE}
\caption{Distribution of contacts between patients and staffs in the hospital.}
\end{table}

\section{Different levels of activity of services}\label{sec:activity}

Figure~\ref{TotalContacts} shows the separation and division of contacts among the 9 services of the hospital, in terms of number of adjacency pairs (a), number of contacts (b) and cumulated length of contacts (c). It reveals some important differences between services. The 5 services including patients seem to be more active than the 4 others, for each of the three criteria. But clear differences also appear between these 5 services.
As one may guess, one reason for this is that services have different sizes (see Figure~\ref{repart-serv}). For adjacency pairs, this is confirmed by the fact that the number of mean adjacency pairs per individual per day varies only a little between two different services (Figure~\ref{TotalContIndiv} (a)). On the other hand, the number of contacts and the cumulated length of contacts per day remain very different from one service to another even when computed in average for one individual (Figure~\ref{TotalContIndiv} (b) and (c)). This indicates that for these two criteria, the sizes of services cannot be hold for entirely responsible of the disparities in the activity of services appearing on Figure~\ref{TotalContacts}.

Services S6 to S9, which do not include any patient, have a mean number of contacts and a mean cumulated length of contacts per individual which is far smaller than those of services S1 to S5, which do include patients (Figure~\ref{TotalContIndiv} (b) and (c)). Moreover, among these latter services, it appears that services S4, S5 and S2 present a higher mean individual activity, for these two parameters, than services S1 and S3; and it turns out that S4, S5 and S2 are the 3 services that contain the greater number of patients (see Figure~\ref{repart-serv}). These observations suggest that the individual activity of patients with regard to number of contacts and cumulated length of contacts may be much higher than the one of staffs.

Another interesting fact revealed by Figures~\ref{TotalContacts} and~\ref{TotalContIndiv} is that the number of contacts and the cumulated length of contacts per service behave very similarly. This is confirmed by Figure~\ref{EquivContLength} (a) which shows the scatter plot of the mean values per individual and per day of these two parameters, for each service of the hospital. The plot shows that for all services, the mean values of number of contacts and cumulated length of contacts are strongly correlated. This is actually an even more general fact as the correlation between these two parameters is also clearly visible for each individual on the whole period of study (Figure~\ref{EquivContLength} (b)). Therefore, as they give very similar results in all the experiments we conducted, we choose to keep only one of them in most of the analysis presented in the rest of the paper, namely the cumulated length of contacts.

\begin{figure}
\centering
\includegraphics[width=\linewidth]{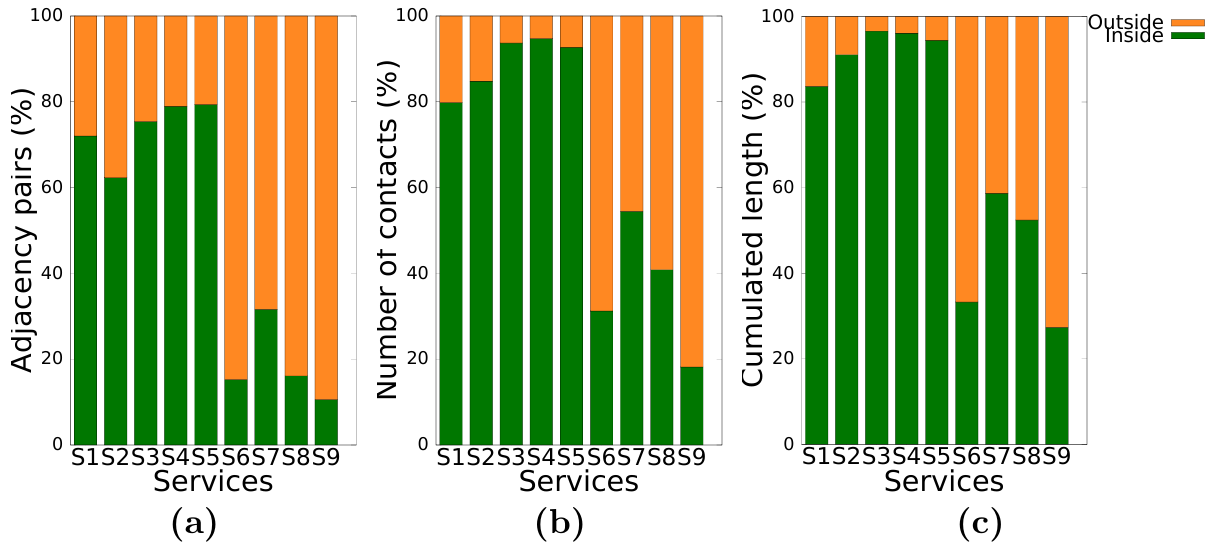}
\caption{Distribution between internal activity and external activity for each service. (a) adjacency pairs, (b) number of contacts, (c) cumulated length of contacts. Internal activity is in dark green and external activity in light orange.}
\label{IntroBrut}
\end{figure}

\begin{figure}
\centering
\includegraphics[width=\linewidth]{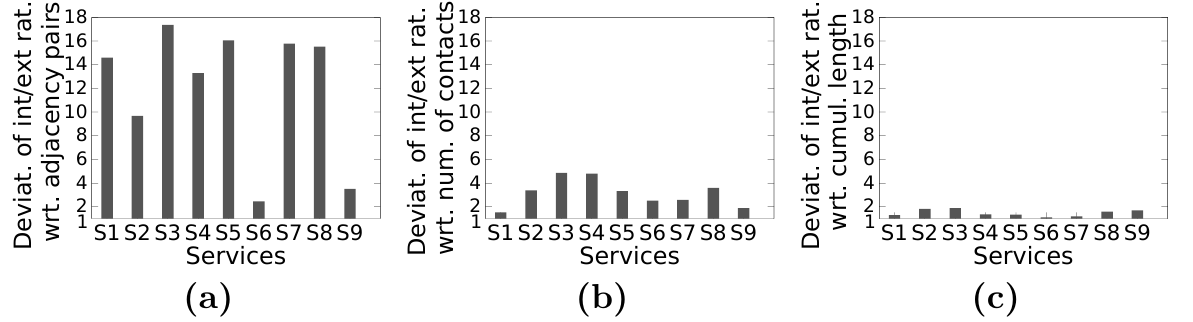}
\caption{Factors of introversion for each service. (a) deviation of the int/ext ratio of adjacency pairs, (b) deviation of the int/ext ratio of number of contacts, knowing adjacency pairs, (c) deviation of the int/ext ratio of cumulated length of contacts, knowing both adjacency pairs and number of contacts.}
\label{DeviaIntrovertion}
\end{figure}

\section{Different behaviours of patients and staffs}\label{sec:PA-ST}

As pointed out above, patients and staffs seem to have a very different activity. We then refine our analysis of the mean activity per individual and per day by separating patients from staffs in the 5 concerned services (Figure~\ref{TotalContactsPAPE}). It turns out that patients are a bit less active than staffs of the same service (about 20\% to 30\% less) in terms of adjacency pairs, but are much more active in terms of cumulated length of contact (between 2 and 6 times more, except for service S1 where cumulated length of contacts of patients and staffs are comparable). This explains why the differences between services that appeared on Figure~\ref{TotalContacts} (a) for the whole service disappear when considering the adjacency pairs per person (Figure~\ref{TotalContIndiv} (a)), while this difference does not disappear for cumulated length (see Figure~\ref{TotalContacts} (c) and Figure~\ref{TotalContIndiv} (c)). Indeed, as the number of adjacency pairs is comparable for patients and staffs, so are the mean individual values for each service, independently of the fact that they contain more patients or more staffs. On the opposite, for cumulated length of contacts, as the one of patients is much higher than the one of staffs, it follows that the mean individual value in services containing both staffs and patients (S2 to S5, S1 being an exception as pointed above) is higher than the mean individual value in services containing only staffs (S6 to S9).

\begin{table}
\centering
\normalsize
\begin{tabular}{c}
\begin{tabular}{c|c|c|c|c}
 &    PA-PA    &PA-ST&    ST-ST&    All \\
 \hline
Ext.&    0.05&    0.23&    0.06&    0.34 \\
Int.&    0.19&    0.33&    0.14&    0.66 \\
\hline\hline
All    &0.24&    0.56&    0.20    &1.00 \\
\end{tabular}
\\[0.6cm]
(a) Global distribution of adjacency pairs
\end{tabular}

\smallskip

\null
\hfill
\begin{tabular}{c}
\begin{tabular}{c|c|c|c}
PA vs &PA    &ST    & All\\
\hline
Ext.    &0.10    &0.22 & 0.32    \\
\hline
Int.    &0.36    &0.32    & 0.68\\
\hline\hline
All    &0.46    &0.54    & 1\\
\end{tabular}
\\[0.6cm]
(b) Patient-centred adj. pairs
\end{tabular}
\hfill
\begin{tabular}{c}
\begin{tabular}{c|c|c|c}
ST vs    &PA    &ST    & All\\
\hline
Ext.   &0.24    &0.12& 0.36    \\
\hline
Int.    &0.34    &0.30 &    0.64\\
\hline\hline
All    &0.58    &0.42&    1\\
\end{tabular}
\\[0.6cm]
(c) Staff-centred adj. pairs
\end{tabular}
\hfill
\null
\caption{Distribution of adjacency pairs between patients and staffs, distinguishing between internal and external pairs. (a) global distribution, (b) patient-centred point of view, (c) staff-centred point of view.\label{IntExt-PAPE}}
\end{table}

These differences rise some crucial questions: what is the role of patients and staffs in the global contact pattern of the hospital? Where is located the majority of contacts? between patients, between staffs or between patients and staffs? Table~\ref{GenStatPAPE} shows that a vast majority of the cumulated length of contacts in the hospital, 80\%, involves two patients, while only 12\% of this length involve one patient and one staff, and 8\% involve two staffs. Nevertheless, the picture for adjacency pairs is quite different: those between patients represent only 24\% of all pairs, which is about 35\% less than in the full-uniform network. The majority of adjacency pairs, 56\%, involves one patient and one staff, and 20\% of them involve two staffs. Both of these values are about 20\% higher than in the full-uniform network. This suggests that the contacts of staffs and in particular the contacts between staffs and patients are very important for the structure of the daily contact network, and may then play a key role regarding the possibility of spreading in the hospital.

Tables~\ref{GenStatPAPE} (b) and (c) give the separation and division of contacts respectively for an average patient and an average staff. They show that the majority of the adjacency pairs of a patient (54\%) occurs with a staff, and that the majority of the adjacency pairs of a staff (58\%) occurs with a patient. Note that, opposite to the case of patients whose cumulated length of contacts is strongly unbalanced in favour of contacts with patients (93\%), staffs share much more equitably their length of contacts between patients (42\%) and staffs (58\%). This confirms that staffs present a more open pattern of contacts than the one of patients, which may result for them in particular spreading abilities.

\begin{table}
\centering
\normalsize
\begin{tabular}{c}
\begin{tabular}{c|c|c|c|c}
&PA PA&PA ST&ST ST&All\\
\hline
Ext.&0.03&0.04&0.01&0.08\\
Int.&0.78&0.07&0.07&0.92\\
\hline\hline
All&0.81&0.11&0.08&1\\
\end{tabular}
\\[0.6cm]
(a) Global distribution of cumulated length
\end{tabular}

\smallskip

\null
\hfill
\begin{tabular}{c}
\begin{tabular}{c|c|c|c}
PA vs&PA&ST&All\\
\hline
Ext.&0.03&0.03&0.06\\
\hline
Int.&0.90&0.04&0.94\\
\hline\hline
All&0.93&0.07&1\\
\end{tabular}
\\[0.6cm]
(b) Patient-centred cum. length
\end{tabular}
\hfill
\begin{tabular}{c}
\begin{tabular}{c|c|c|c}
ST vs&PA&ST&All\\
\hline
Ext.&0.16&0.08&0.24\\
\hline
Int.&0.26&0.50&0.76\\
\hline\hline
All&0.42&0.58&1\\
\end{tabular}
\\[0.6cm]
(c) Staff-centred cum. length
\end{tabular}
\hfill
\null
\caption{Distribution of cumulated length of contacts between patients and staffs, distinguishing between internal and external contacts. (a) global distribution, (b) patient-centred point of view, (c) staff-centred point of view.\label{IntExt-PAPE-length}}
\end{table}

\section{Introversion and interconnection of services}\label{sec:introvertion}

We mentioned previously that most of the activity of the link stream takes place inside services. Here we investigate further this question by examining the introversion of each service with regard to adjacency pairs, number of contacts and cumulated length of contacts. In all the rest of the article, we qualify contacts and adjacency pairs as \emph{internal} or \emph{external} depending on whether they take place inside a service or between two distinct services. Figure~\ref{IntroBrut} gives the separation and division for each service between its internal and external activity, in terms of adjacency pairs (a), number of contacts (b) and cumulated length of contacts (c). One can clearly distinguish two groups of services on the three plots: services S1 to S5 which have the bigger part of their activity occurring inside the service itself and services S6 to S9 which have the bigger part of their activity occurring outside the service. Plot (a) shows that more than 60\% of the adjacency pairs of services S1 to S5 are internal. For other services, the situation appears to be reversed: S6, S8 and S9 have less than 20\% of internal adjacency pairs and S7 less than 35\%. For number of contacts (b) and cumulated length of contacts (c), for all services the proportion of internal activity is augmented compared to the proportion of internal adjacency pairs. For services S1 to S5, all of these values are above 80\%, while they are between 18\% and 53\% only for services S6 to S9. These differences are partly explained by the fact that services S6 to S7 contains only staffs, including many healthcare workers, but no patients. It is then natural that a large proportion of the contacts of the members of these services occur outside the service, for example with individuals from the services containing patients. But there are other factors that may explain the differences observed between services S1 to S5 and services S6 to S9 in the separation and division between their internal and external activity. The size of the service for example have a strong impact on this division, as smaller services are likely to have a more important part of their activity directed outside of the service. One way to separate the contribution of the size of services in our analysis, and then isolate the contribution of the functional characteristics of services, is to introduce the notion of \emph{factor of introversion}.

Formally, for $\alpha$ being one of the three parameters we use (namely adjacency pairs, number of contacts and cumulated length of contacts), the \emph{int/ext ratio} of a service $S$ with regard to $\alpha$ is defined as $\alpha_{int}(S)/\alpha_{ext}(S)$, where $\alpha_{int}(S)$ is the value of parameter $\alpha$ (e.g. number of adjacency pairs) inside $S$ and $\alpha_{ext}(S)$ is the value of parameter $\alpha$ between $S$ and the rest of the hospital. Then, the \emph{factor of deviation of the int/ext ratio} of service $S$, which we also call \emph{factor of introversion}, is defined as the quotient between the int/ext ratio of $S$ in the real aggregated network and the int/ext ratio of $S$ in some specifically defined uniform network. For adjacency pairs, we use for comparison the full-uniform network (cf. Definition~\ref{def:fulluniform}). For number of contacts, we use the \emph{contact-uniform network}, which has exactly the same adjacency pairs as the real aggregated network, and for cumulated length of contacts we use the \emph{length-uniform network}, which has the same adjacency pairs as the real aggregated network, each of which has the same number of contacts as in the real aggregated network.

\begin{figure}
\centering
\includegraphics[width=\linewidth]{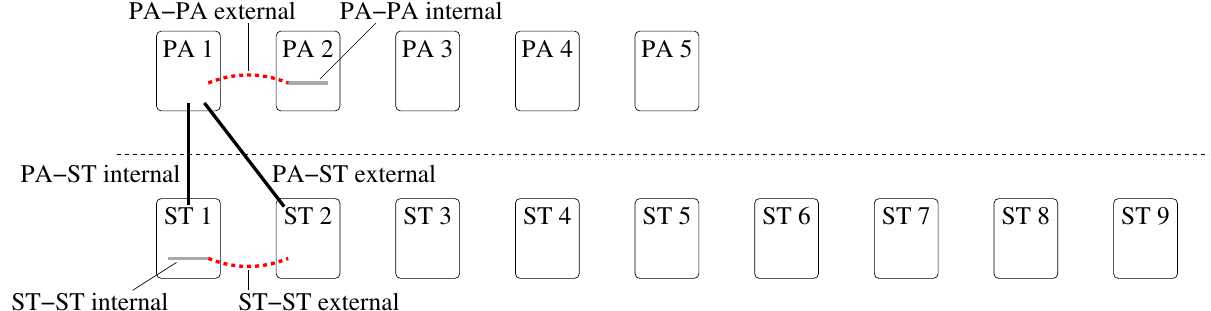}
\caption{Bipartite-like structure of the aggregated network between classes of patients and classes of staffs. One class is constituted either by the patients (up) or by the staffs (bottom) of one single service. Black bold lines are for links crossing the bipartition, grey lines for those inside one class and red doted lines are for the defaults of the bipartition, i.e. links between two distinct classes of patients or two distinct classes of staffs.}
\label{Biparti}
\end{figure}			

\begin{definition}[Contact-uniform network]
The contact-uniform network associated to a link stream $L$ is the graph $G=(V(L),E(L))$ made of the adjacency pairs of $L$, where each adjacency pair $\set{u,v}$ is assigned to the same number of contacts $\#cont(u,v)=\#cont(L)/\#pairs(L)$, which is the mean number of contacts per adjacency pairs in $L$.
\end{definition}

\begin{definition}[Length-uniform network]
The length-uniform network associated to a link stream $L$ is the graph $G=(V(L),E(L))$ made of the adjacency pairs of $L$, where each adjacency pair $\set{u,v}$ is assigned to its number $\#cont_{\set{u,v}}$ of contacts in $L$ and a cumulated length of contact $cumul\_length(u,v)=\#cont_{\set{u,v}}$ $.(cumul\_length(L)/\#cont(L))$, which is the cumulated length obtained for $\set{u,v}$ by making all contact lengths equal to the mean length of contacts of $L$.
\end{definition}

The rational behind these definitions is that for the number of contacts, we compute its deviation knowing the adjacency pairs of the real aggregated network, and for the cumulated length of contacts, we compute its deviation knowing both the adjacency pairs and the number of contacts of the real aggregated network. This allows to test whether the deviation observed for one parameter is a consequence of those observed for other parameters.

The results are depicted on Figure~\ref{DeviaIntrovertion}. They confirm, as observed on Figure~\ref{IntroBrut}, that services S1 to S5 are strongly introverted in terms of adjacency pairs: they have a factor of introversion between 9 and 18. More surprisingly, services S7 and S8 also appear to be strongly introverted, which was not predictable from Figure~\ref{IntroBrut}. Their factor of introversion are both above 15, meaning that the ratio of adjacency pairs between inside and outside the service is 15 times higher (in favour of internal pairs) that what it would be if contacts between individuals in the hospital occurred completely freely, independently of spatial, organisational and functional constraints. Even services S6 and S9, which do not have a single determined location in the building of the hospital, also appear to be more introverted than expected in the full-uniform network, their factor of introversion being higher than 2. Then, despite of what we could expect from Figure~\ref{IntroBrut}, all services strongly favour adjacency pairs inside the service rather than outside, in a very unbalanced way for at least 7 out of 9 of them which have a factor of deviation higher than 9.

\begin{figure}
\centering
\includegraphics{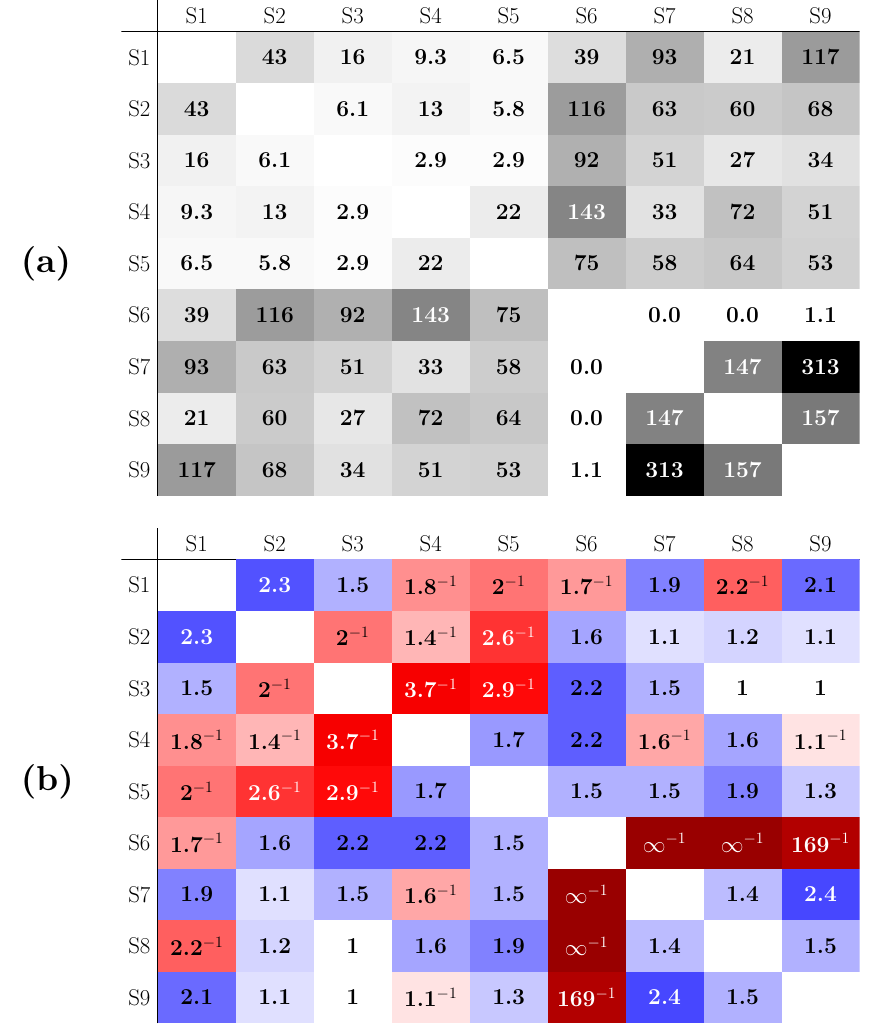}
\caption{Affinities between services based on adjacency pairs. (a) mean density (in thousandths) of adjacency pairs per day between pairs of distinct services, (b) factors of deviation of these values. We use exponent $^{-1}$ to denote the inverse of a number, e.g. $2.6^{-1}$ means $1/2.6$, and $\infty$ stands for an infinite value ($\infty^{-1}$ stands for $1/\infty=0$). The blue scale of colours is for favoured relationships and the red one for unfavoured relationships. Both matrices are symmetric.}
\label{PairsBetServ}
\end{figure}			

Going further, even knowing this unbalanced structure of the adjacency pairs, services are still clearly introverted in terms of number of contacts (factors between 1.5 and 5). This means that services do not have only a strong preference for making adjacency pairs inside rather than outside, but they are also much more likely to repeat contacts for their internal adjacency pairs. This trend seems to be rather equitably shared by all services without strong difference between the group of services including patients and staffs (S1 to S5) and the one of services that include only staffs (S6 to S9).
For cumulated length, the factor of introversion is less than 2 for all services, but always strictly greater than 1. The fact that these values are lower than the previous ones is a consequence of the correlation between cumulated length of contacts and number of contacts (see Figure~\ref{EquivContLength}). But still, they indicate that services not only favour internal adjacency pairs and internal repetition of contacts, but also prefer longer contacts between their members rather than outside. Again, this trend is of comparable strength in all the services.

\begin{figure}
\centering
\includegraphics{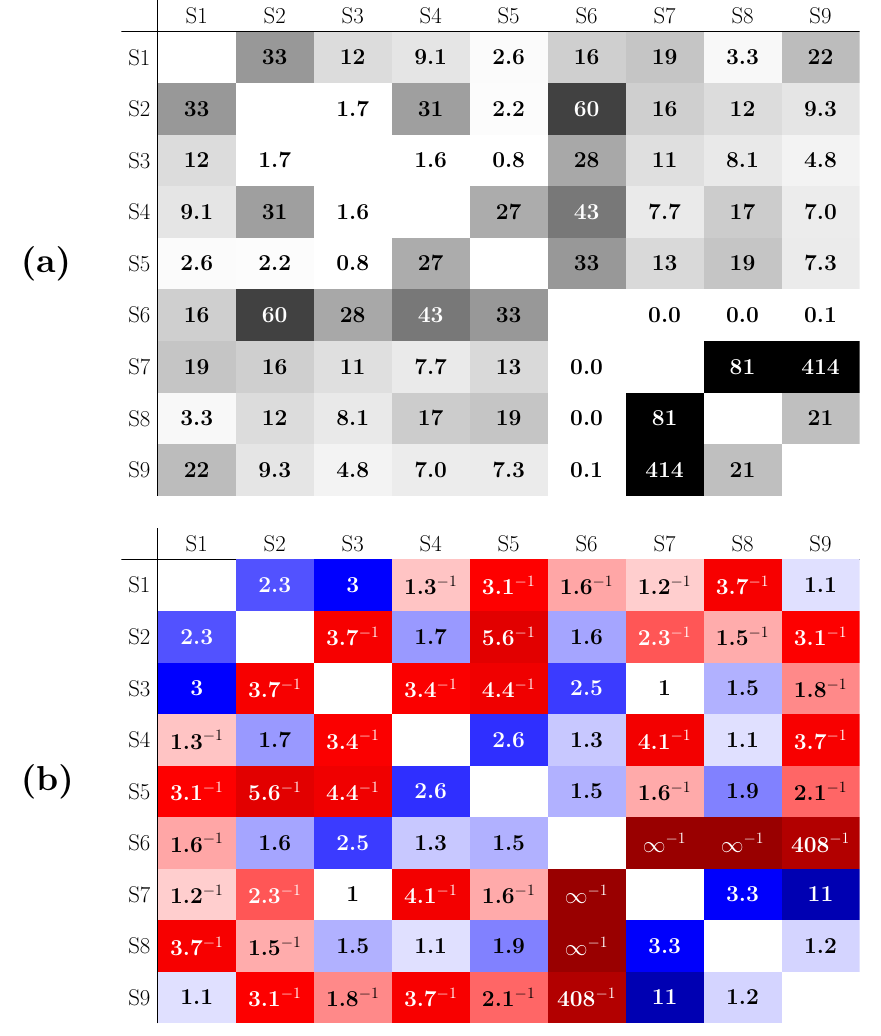}
\caption{Affinities between services based on cumulated length of contacts. (a) mean cumulated length of contacts (in seconds) per pair of individuals and per day between pairs of distinct services, (b) factors of deviation of these values. We use exponent $^{-1}$ to denote the inverse of a number, e.g. $2.6^{-1}$ means $1/2.6$, and $\infty$ stands for an infinite value. The blue scale of colours is for favoured relationships and the red one for unfavoured relationships. Both matrices are symmetric.}
\label{LengthBetServ}
\end{figure}			

Table~\ref{IntExt-PAPE} gives some global statistics distinguishing both between internal and external adjacency pairs and between patients and staffs. It reveals a strong bipartite-like structure of the aggregated network between the staffs divided into services on one side (9 classes), and the patients divided into services on the other side (5 classes), see Figure~\ref{Biparti}. Indeed, more than 83\% (=0.56/(0.56+0.05+0.06)) of the adjacency pairs between these 14 classes occur between one patient and one staff. In addition, links between patients and staffs represent more than 67\% of the external links between services of the hospital (18\% of these links occur between staffs and 15\% between patients). This shows that the contacts between patients and staffs play a prevalent role in connecting the introverted services of the hospital. These observations are confirmed from an individual centred point of view (see Table~\ref{IntExt-PAPE} (b) and (c)): an individual (either patient or staff) has only few external adjacency pairs with his own side of the bipartition, while the distribution between its external and internal pairs with the other side are more balanced than internal/external pairs in the whole hospital.

\begin{figure}
\centering
\includegraphics{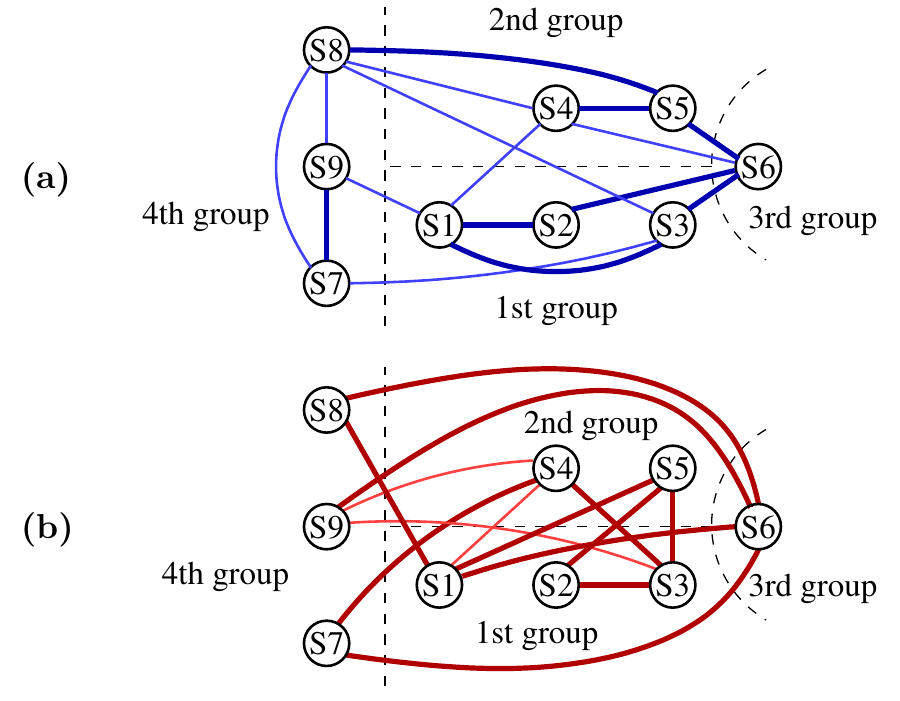}
\caption{Structure of the relationships between services of the hospital. (a) favoured relationships, in blue, (b) unfavoured relationships, in red. Only clearly favoured or unfavoured relationships (thin light-coloured lines) and strongly favoured or unfavoured relationships (bold dark-coloured lines) are depicted.}
\label{Services}
\end{figure}			

Table~\ref{IntExt-PAPE-length} gives the same kind of statistics as Table~\ref{IntExt-PAPE}, but considering cumulated length of contacts instead of adjacency pairs. The resulting picture of the hospital is quite different. Firstly, 78\% of the total cumulated length of contacts in the hospital occurs between two patients in a same service, which was not at all the case for adjacency pairs. We give more explanation about this fact in Section~\ref{sec:temp}, dedicated to the temporal structure of contacts, by considering the times of the day when these contacts occur. Secondly, patients also play a more important role in the cumulated length of contacts between two different services: 87\% of this length is made by contacts involving at least one patient and this proportion is shared in a balanced way between contacts involving two patients and contacts involving one patient and one staff. The importance of cumulated length of contacts involving patients for connecting different services is confirmed by the patient centred view (Table~\ref{IntExt-PAPE-length} (b)) and the staff centred view (Table~\ref{IntExt-PAPE-length} (c)). Omitting the time one patient spends with patients of his/her service, the rest of his cumulated length of contacts is very equitably shared between staffs of his/her own service, staffs of other services and patients of other services. A staff spends a large proportion of its cumulated length of contacts with staffs of his service (50\%) and only few with staffs of other services (8\%). Staffs also clearly favour cumulated length of contacts with patients inside their service rather than outside. This shows, that opposite to the case of adjacency pairs, where pairs involving staffs seem particularly important in connecting the different services between them, for cumulated length of contacts, it is the contacts of patients that seem to be prevalent between different services. This probably results in qualitatively different abilities of patients and staffs for spreading diffusions in the hospital.

\section{Affinities between services}\label{sec:acquaintance}

\begin{figure}
\centering
\includegraphics{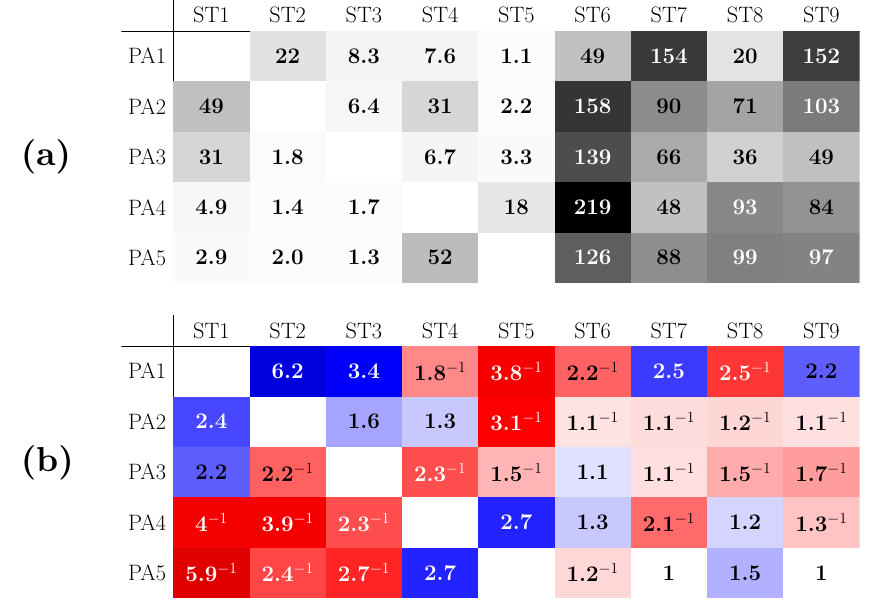}
\caption{Affinities between classes of patients and staffs based on adjacency pairs. (a) mean density (in thousandths) of adjacency pairs per day between classes of patients (vertical) and classes of staffs (horizontal) in distinct services, (b) factors of deviation of these values. We use exponent $^{-1}$ to denote the inverse of a number, e.g. $2.6^{-1}$ means $1/2.6$. The blue scale of colours is for favoured relationships and the red one for unfavoured relationships.}
\label{PairsPAPEServ}
\end{figure}

The question we address in this section is to determine whether some pairs of services are more likely to interact between them than others. Figure~\ref{PairsBetServ} (a) gives the average density of adjacency pairs per day between every pair of services $S_i S_j$, i.e. the number of adjacency pairs in one day between $S_i$ and $S_j$ divided by the number of possible pairs on this day $|S_i||S_j|$. Similarly, Figure~\ref{LengthBetServ} (a) gives the average cumulated length of contacts per day and per pair of individuals between $S_i$ and $S_j$. The rational for dividing by the number of possible pairs between services $S_i$ and $S_j$ is that the values obtained describe the intensity of interactions between service $S_i$ and $S_j$, independently of their size.
As we have shown earlier, all services do not have the same activity level. Naturally, two services $S_i$ and $S_j$ having higher activity levels will tend to have a higher intensity of interactions between them. This does not mean that $S_i$ and $S_j$ favour the interactions between them compared to interactions with other services: this higher intensity simply results from the fact that both of them interact more with everyone in the hospital. With the aim of constraining or rearranging contacts inside the hospital for limiting the possible diffusions of bacteriological strains, it is highly desirable to be able to identify which pairs of services favour the interactions between them compared to their interactions with other services. This is precisely the goal of the deviation factors given on Figure~\ref{PairsBetServ} (b) and Figure~\ref{LengthBetServ} (b). Formally, the deviation factor is defined as usual as the ratio between the value (here the intensity of interactions, appreciated either by adjacency pairs or by cumulated length of contacts) in the real aggregated network and in a random network used for comparison, here the \emph{configuration model} without internal contacts inside services (the diagonal of all matrices is empty). In the configuration model~\cite{MR95}, each service is assigned to the same number of semi-pairs as in the real aggregated network, and each of these semi-pairs is randomly matched with another semi-pair, here, of another service, as we do not consider internal contacts. If we denote $|D_i|$ the number of semi-pairs assigned to $S_i$, in average, the expected number of pairs between $S_i$ and $S_j$ resulting from this random matching process is $\frac{|D_i||D_j|}{|D|}$, where $|D|$ is the total number of semi-pairs in the whole hospital. One can proceed similarly with duration of contacts instead of adjacency pairs. In this case, the expected cumulated length of contacts between $S_i$ and $S_j$ is $\frac{|E_i||E_j|}{|E|}$, denoting $|E_i|$ the cumulated length of semi-contacts of $S_i$ and $|E|$ the total cumulated length of all semi-contacts in the hospital. This average network is precisely the one we use for computing deviations. The meaning of the deviation factor is to show which pairs of services have more (deviation factor greater than $1$) or less (deviation factor less than $1$) interactions between them that what is expected from their respective activity levels. When the deviation factor between $S_i$ and $S_j$ is greater than $1$, we say that the relationship between $S_i$ and $S_j$ is favoured and when their deviation factor is less than $1$ we say that their relationship is unfavoured.

\begin{figure}
\centering
\includegraphics{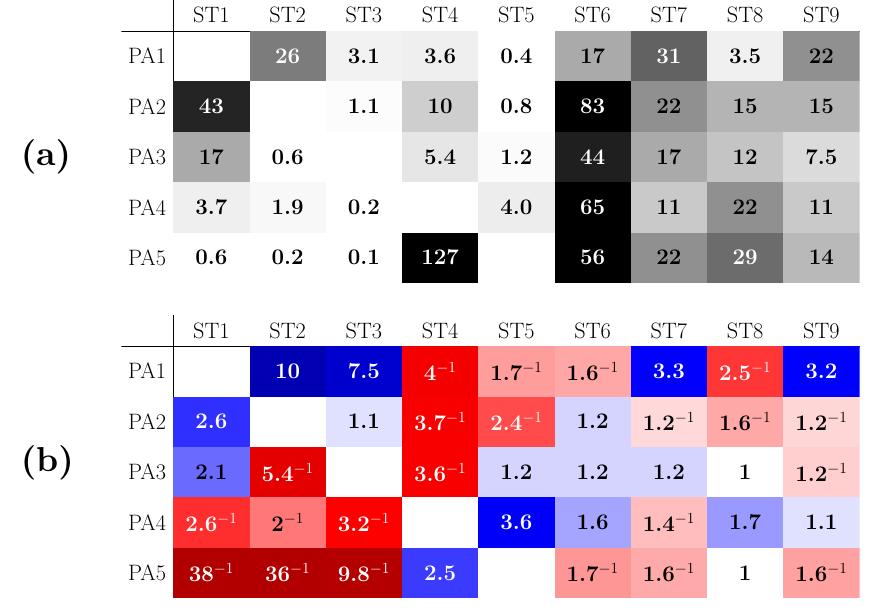}
\caption{Affinities between classes of patients and staffs based on cumulated length of contacts. (a) mean cumulated length of contacts (in seconds) per pair of individuals and per day between classes of patients (vertical) and classes of staffs (horizontal) in distinct services, (b) factors of deviation of these values. We use exponent $^{-1}$ to denote the inverse of a number, e.g. $2.6^{-1}$ means $1/2.6$. The blue scale of colours is for favoured relationships and the red one for unfavoured relationships.}
\label{LengthPAPEServ}
\end{figure}

Looking at the statistics for adjacency pairs (Figure~\ref{PairsBetServ} (a)), it appears that the large majority of adjacency pairs occurs between individuals of services S1 to S5 on one side, which are the services containing both patients and staffs, and individuals of services S6 to S9 on the other side, which contain only staffs. Looking closer, and taking into account the deviations of these values (Figure~\ref{PairsBetServ} (b)), one can actually distinguish finer groups of services. The first two groups are S1, S2, S3 and S4, S5. While the interactions inside each of these groups are high, it turns out that the deviation factors between the two groups have high unfavoured values, meaning that services S1, S2, S3 have a strong tendency not to interact with S4 and S5. This probably comes from the fact that each of these two groups of services occupies a distinct wing of the building, which are physically separated (see map in Figure~\ref{fig:plan}). The third group contains only service S6 which interacts a lot with services S1 to S5 but has almost no interactions with services S7 to S9. The reason is that S6 is the night service and therefore involves staffs working in other services of the hospital, except the reeducation services (S7 to S9) which work only at daytime. Finally, services S7 to S9, strongly connected together, constitute the fourth group.

\begin{table}
\normalsize
\centering
\begin{tabular}{l|l|}
\hline
C1 & assistant nurses \\
C2 & nurses \\ 
C3 & housekeepers \\
C4 & intern nurses \\
C5 & manager nurses\\ 
C6 & stretcher-bearers \\
\hline
\end{tabular}
\begin{tabular}{l|l|}
\hline
C7 & physiotherapists \\
C8 & occupational therapists \\
C9 & other reeducators \\
C10 & organiser and hairdresser \\
C11 & logistic \\
C12 & administration \\
\hline
\end{tabular}
\label{tab:categories}
\caption{Titles of the twelve socio-professional categories of the hospital.}
\end{table}

\begin{figure}
\centering
\includegraphics{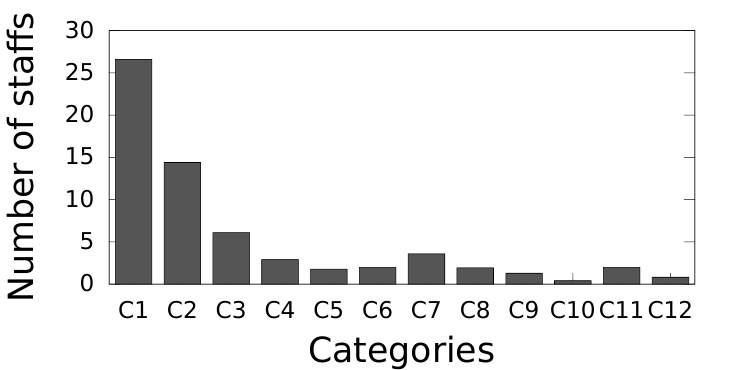}
\caption{Mean number of staffs per day in each socio-professional category.}
\label{fig:RepartCat}
\end{figure}

The picture given by the cumulated length of contacts is similar, with some meaningful differences. Note that the deviations for cumulated length of contacts are often higher than the one observed for adjacency pairs: the favoured and unfavoured relationships are more clearly marked when seen through duration of contacts. On the other hand, the general organisation into groups highlighted above through adjacency pairs is still visible on cumulated length of contacts. The main difference is that more than half of the relationships between services S1 to S5 and services S6 to S7 appear to be unfavoured in terms of cumulated length of contacts while they were almost all favoured in terms of adjacency pairs, revealing a more complex pattern of contacts in the hospital.

In order to get a deeper insight into the structure of the relationships between services of the hospital, we used both adjacency pairs and cumulated length of contacts to build the graph of favoured relationships between services on Figure~\ref{Services} (a). In the drawing, only pairs of services $S_i S_j$ that have a \emph{clearly favoured} relationship are linked by a line, i.e. both deviation factors, for adjacency pairs and cumulated length, between $S_i$ and $S_j$ are greater than $1$ and at least one of them is greater than $1.5$. Moreover, if both deviation factors are greater than $1.5$, then the line linking $S_i$ and $S_j$ on the drawing is thicker and darker and we say that the relationship is \emph{strongly favoured}. One can retrieve on Figure~\ref{Services} (a) the details of the big picture of relationships between services described above. Similarly, Figure~\ref{Services} (b) shows the graph of unfavoured relationships between services. It is remarkable that the relationships between services of the hospital are strongly polarised: 29 pairs of services out of 36 possible pairs are either clearly favoured (15 of them, and 8 of them are even strongly favoured) or clearly unfavoured (14 of them, and 11 of them are strongly unfavoured). Moreover, looking closer, only 3 out of the 7 remaining pairs (i.e. about 8\% of the total number of possible pairs) have a balanced deviation on the favoured and on the unfavoured sides (for adjacency pairs on one side and for cumulated length of contacts on the other side) that forbids to mark them as favoured or unfavoured. The 4 others are strongly deviated on one side and only slightly deviated on the other side. These pairs may then safely be classified as favoured (such as S7S1) or as unfavoured (such as S7S2, S9S2, S9S5). Totally, more than 90\% of the relationships between services of the hospital are polarised (and more than 80\% are clearly polarised in the sense above). This shows that taking into account the structure of relationships between services in the analysis of the diffusions occurring in the hospital is certainly relevant.

\begin{figure}
\centering
\includegraphics[width=\linewidth]{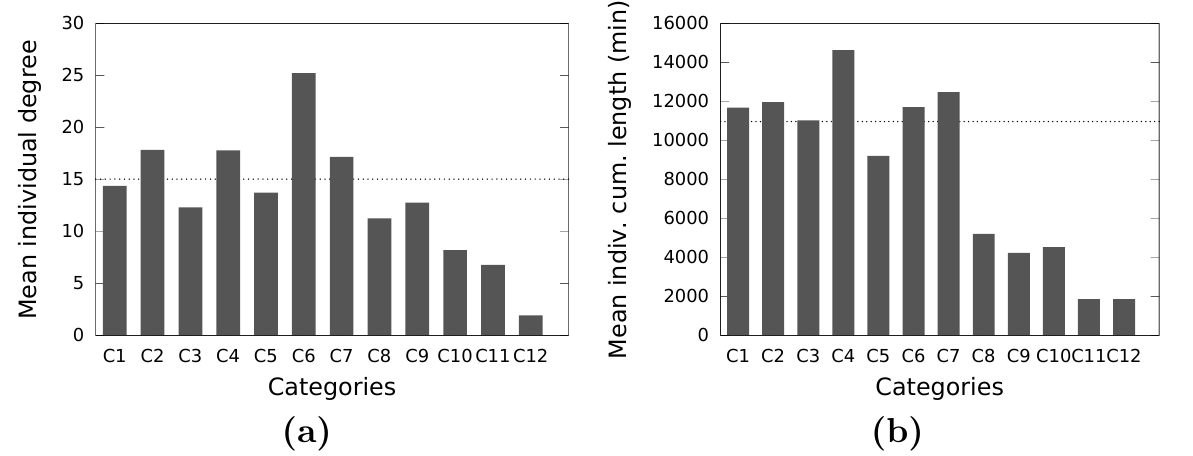}
\caption{Mean activity per staff and per day for each category. (a) number of adjacency pairs, (b) cumulated length of contacts. This includes contacts between staffs and patients. The doted lines depict the mean values per staff in the whole hospital.}
\label{TotalContactsPECat}
\end{figure}

As we pointed out earlier, services are widely introverted and the connections between them strongly relies on contacts between staffs and patients. In order to get a deeper insight into the structure of interconnections of services through contacts between patients and staffs, we apply the methodology above restricted to these contacts. Figure~\ref{PairsPAPEServ} gives the intensity of relationships and their deviation factors for adjacency pairs and Figure~\ref{LengthPAPEServ} for the cumulated length of contacts.

We retrieve the big picture revealed above (see Figures~\ref{PairsBetServ} and~\ref{LengthBetServ}) with some accurate details. 
The relationships between services S1 to S5 restricted to interactions between patients and staffs only appear pretty similar to those observed at the level of whole services, except the clearly favoured relationship between patients of S2 and staffs of S3 which was not visible before. In general, the relationships between services S1 to S5 are more clearly marked, both favourably or unfavourably, on contacts between patients and staffs. It is also remarkable that almost all these relationships are symmetric (except the one between S2 and S3): if patients of $S_i$ have a clearly favoured relationship with staffs of $S_j$ then so have the staffs of $S_i$ with the patients of $S_j$. Nevertheless, these clearly favoured relationships may be quite unbalanced (as for S1S2 and S1S3): patients of $S_i$ favour contacts with staffs of $S_j$ much more than staffs of $S_i$ do with patients of $S_j$.

Regarding the relationships between services S6 to S9 and services S1 to S5, the pattern revealed by considering only contacts between patients and staffs is slightly different than the one obtained at the level of whole services. The main difference is for the night service S6. Most of its clearly favoured relationships with services S2 to S5 disappear when considering only contacts with patients of S2 to S5. Only its relationship with S4 remains clearly favoured. This shows that an important part of the contacts of the night service occurs with staffs of services S1 to S5 and fewer with patients.
The pattern of contacts between reeducation services (S7 to S9) and healthcare services is slightly changed when considering only patient-staff contacts: S8 looses its clearly favoured relationship with S3 and the one of S7 with S3 is moved to S1. Moreover, the strength of these favoured or unfavoured relationships (see Figures~\ref{PairsPAPEServ} (b) and~\ref{LengthPAPEServ} (b)) are higher for patient-staff interactions than what they were at the level of the whole services. This confirms our previous observation on the importance of contacts between patients and staffs and shows that these contacts strongly shape the structure of contacts in the hospital.

\section{Organisation of the hospital with regard to socio-professional categories}\label{sec:cat}

\begin{sloppypar}
In this section, we investigate the structure of contacts between socio-professional categories of staffs. To this purpose, we apply to the staffs divided into socio-professional categories the same methodology we used for the whole hospital divided into services. The specific organisation of the contacts of staffs is of high interest for epidemiological issues. Firstly, the structure of these contacts is deeply constrained by the role of each category of staffs and this strongly shapes the general picture of all the contacts in the hospital. Secondly, contacts of staffs are the more likely to be changed by changing the organisation and policy of the hospital, while only limited constraints can be applied to patients.
\end{sloppypar}

In the hospital of Berck-sur-mer, staffs are divided into 12 different professional categories. Their titles are given in Table~\ref{tab:categories} and their sizes, in average number of individuals per day, are given on Figure~\ref{fig:RepartCat}. It shows that most of the categories have very few representatives a day (at most 4) except categories C1 to C3, which concentrate a large majority of the staffs of the hospital.
We can a priori distinguish three groups of socio-professional categories which have similar functions in the hospital. The first group is made of categories C1 to C5, which are nurses and housekeepers. These categories are the more present and visible in the daily life of the hospital and potentially constantly working in contact with other people in the hospital, especially patients. The second functional group is the one made of categories C7 to C9 which contain reeducation staffs. These categories are specialised and give more occasional cares to patients. Finally, categories C11 and C12 are logistic and administrative staffs of the hospital. They are less mobile and less likely to be in contact with patients. Categories C6 (stretcher-bearers) and C10 (organisers and hairdressers) have roles that do not fit directly into the functional groups given above. Nevertheless, their type of activity, devoted to specific and occasional services to patients, makes them closer to the group of reeducation staffs. The order we chose for categories, which is the same in all the figures below, respects the functional groups identified above and is at the same time adapted to the groups identified by our subsequent analyses.

\begin{figure}
\centering
\includegraphics{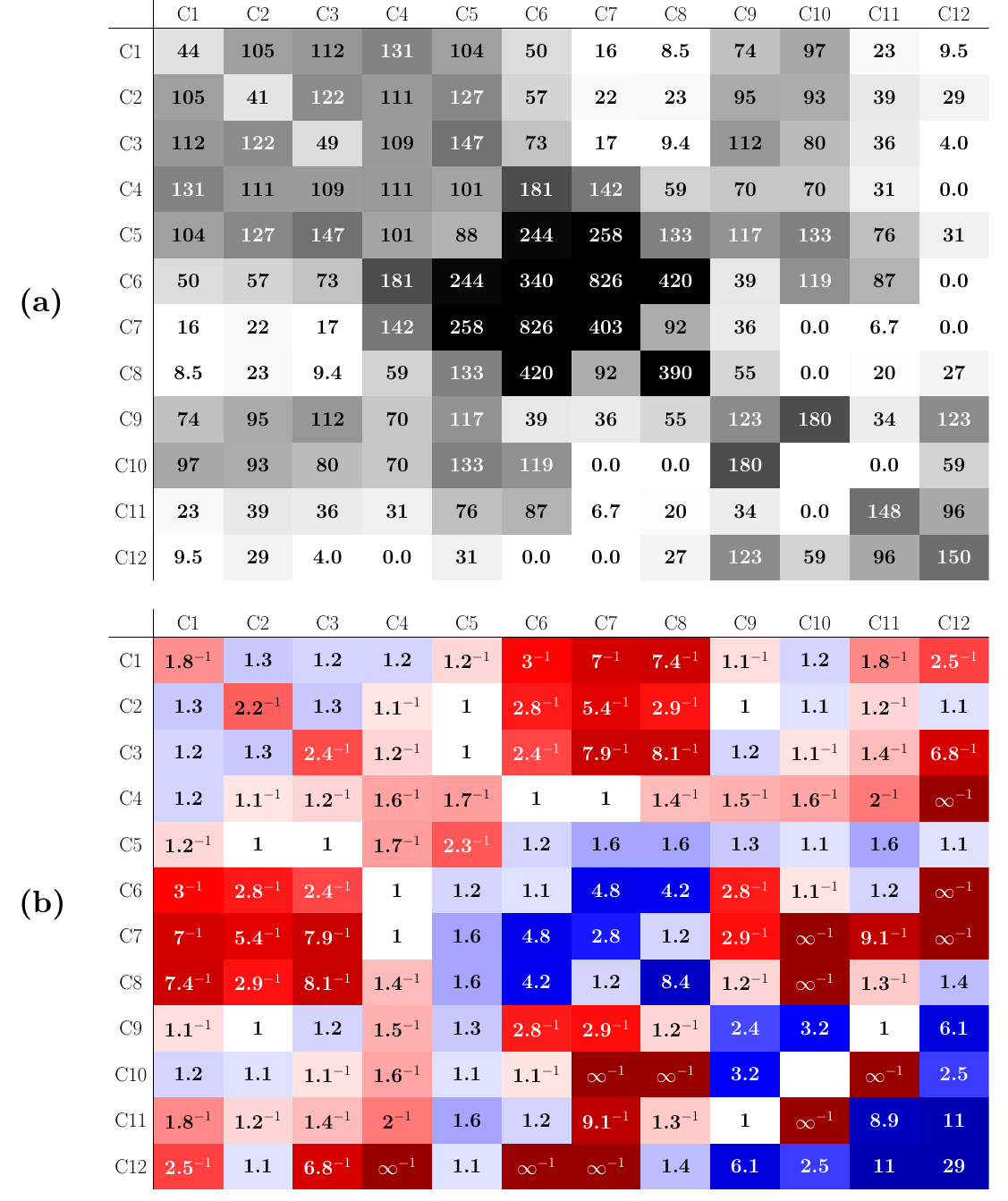}
\caption{Affinities between categories based on adjacency pairs. (a) mean density (in thousandths) of adjacency pairs per day between pairs of categories, (b) factors of deviation of these values. We use exponent $^{-1}$ to denote the inverse of a number, e.g. $2.6^{-1}$ means $1/2.6$, and $\infty$ stands for an infinite value. The blue scale of colours is for favoured relationships and the red one for unfavoured relationships. Both matrices are symmetric.}
\label{fig:adj-cat}
\end{figure}

Figure~\ref{TotalContactsPECat} shows the average daily activity of one staff in each category, in terms of number of adjacency pairs (a) and cumulated length of contacts (b). It takes into account both the contacts between two staffs and the contacts between one staff and one patient.
Both plots reveal two groups of categories having different level of activity. In Figure~\ref{TotalContactsPECat} (a), categories C1 to C9 appear to have a higher and rather similar number of adjacency pairs, while this number is relatively lower for categories C10 to C12. For cumulated length of contacts, the two groups are slightly different than those for adjacency pairs, and the difference between them is much more clearly marked: staffs in categories C1 to C7 have a much longer cumulated length of contacts than staffs in categories C8 to C12. Staffs of C11 and C12, logistic and administration, can even be distinguished from the rest of this group by their particularly short daily cumulated length of contacts. In summary, both adjacency pairs and cumulated length of contacts reveal that staffs whose function is devoted to daily care or rehabilitation of patients have a higher level of activity than other categories, therefore implying a different exposition to diffusion for socio-professional categories.

\begin{figure}
\centering
\includegraphics{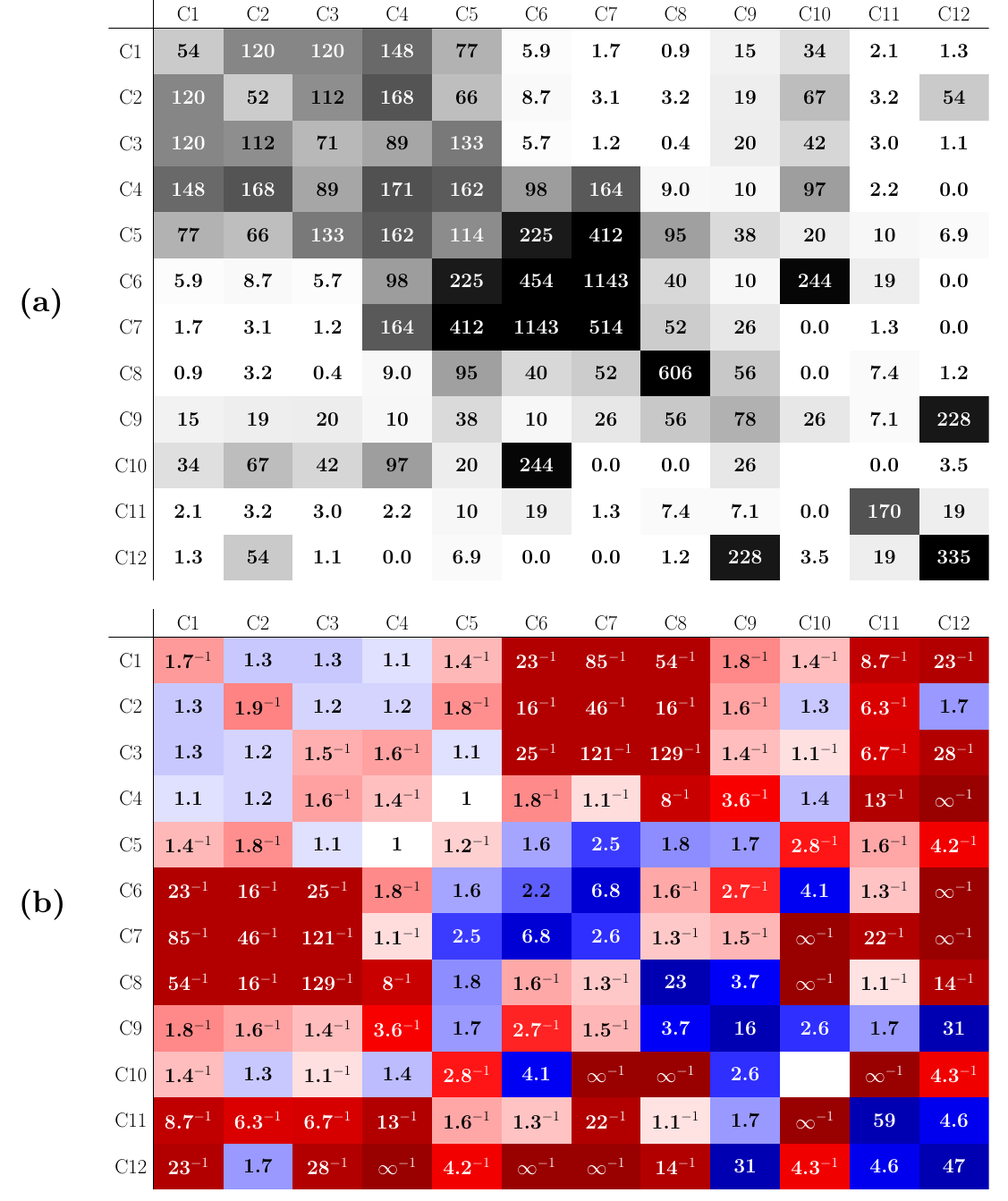}
\caption{Affinities between categories based on cumulated length of contacts. (a) mean cumulated length of contacts (in seconds) per pair of staffs and per day between pairs of categories, (b) factors of deviation of these values. We use exponent $^{-1}$ to denote the inverse of a number, e.g. $2.6^{-1}$ means $1/2.6$, and $\infty$ stands for an infinite value. The blue scale of colours is for favoured relationships and the red one for unfavoured relationships. Both matrices are symmetric.}
\label{fig:length-cat}
\end{figure}

As done for services in Section~\ref{sec:acquaintance}, Figure~\ref{fig:adj-cat} (a) gives the average density of adjacency pairs per day between every pair of categories $C_i C_j$, i.e. the number of adjacency pairs in one day between $C_i$ and $C_j$ divided by the number of possible pairs on this day $|C_i||C_j|$. Figure~\ref{fig:adj-cat} (b) gives the deviation of these values compared to the configuration model. Figures~\ref{fig:length-cat} (a) and (b) give the same for cumulated length of contacts. As usual, the density shows where the interactions are more intense, and its deviation shows which interactions are favoured, i.e. the affinities between categories.

The big pictures arising from the observation of these statistics for adjacency pairs and for cumulated length of contacts are essentially the same, though it appears more clearly for cumulated length of contacts. They reveal a quite specific structure of contacts between socio-professional categories as the contacts between staffs are not at all equitably shared between categories. The most intense relationships as well as the most favoured relationships actually occur within three (overlapping) groups of categories: the first group formed by categories C1 to C5, the second group by categories C5 to C8, and the third one by categories C9 to C12. This partition of the categories in three groups is very clearly marked. It is emphasized by the fact that categories C1 to C3 have strongly unfavoured relationships with categories C6 to C8 (see Figure~\ref{fig:adj-cat} (b) and Figure~\ref{fig:length-cat} (b)) and by the fact that categories C11 and C12 have unfavoured relationships with most of the other categories (see Figure~\ref{fig:length-cat} (b)), except C9. This group structure is articulated by C5, which belongs to both the first and second group (with a preference for the second one), and by the couple C8, C9, which have a strong affinity in terms of cumulated length of contacts (see Figure~\ref{fig:length-cat} (b)) and which then create a bridge between the second and the third group. The particular position of C9 is even accentuated by the fact that it is strongly tied with the rest of the third group.
Outbound of this partition into three groups, C9 and C10 also plays a transversal role by having some affinities with categories belonging to other groups, both for adjacency pairs and for cumulated length of contacts. This structure therefore confers key roles to categories C5, C8, C9 and C10, whose impact on diffusion properties of the link stream of contacts in the whole hospital is worth investigating. Finally, another interesting fact revealed by these analyses is that categories of the first group, C1 to C5, are the only ones that unfavour contacts within their own category, both for adjacency pairs and for cumulated length of contacts. The impact of this on the diffusions occurring into this group should also be deeper investigated.

\begin{figure}
\centering
\includegraphics{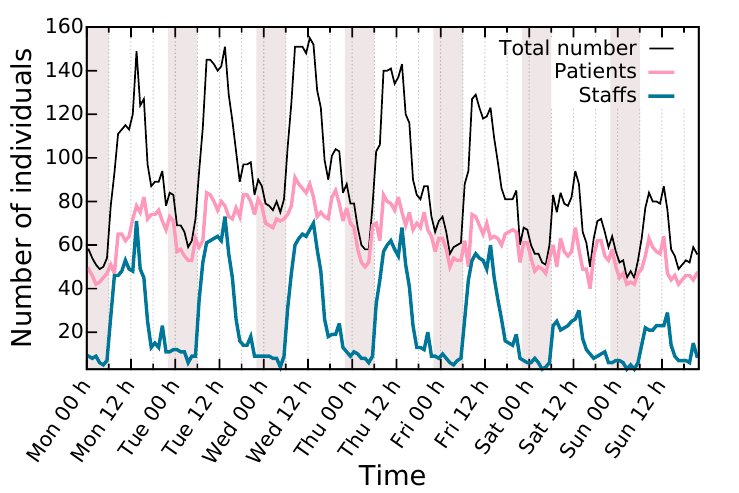}
\caption{Number of active individuals per hour in the hospital. Patients are in light pink, staffs in dark turquoise and the sum of all individuals is depicted by a thin black line.}
\label{temp-pattern-a}
\end{figure}

\section{Temporal structure of contacts in the hospital}\label{sec:temp}

In this section we study the evolution over time of contacts in the hospital. Figure~\ref{temp-pattern-a} gives the evolution over one week (from Monday July 6th to Sunday July 12th) of the number of active individuals per hour, i.e. the individuals who had at least one contact during the considered hour. One retrieves the circadian rhythm of many human activities: the number of active individuals is higher between 6:00 and 24:00, with a peak around 12:00, when people gather for lunch, and this number is much lower between 0:00 and 6:00. There is also a clear weekly pattern denoted by a lower number of active individuals during week-ends.
Figure~\ref{temp-pattern-a} also distinguishes between the number of active patients and the number of active staffs. Both of these numbers observe the same circadian and weekly pattern. Nevertheless, the variation of the number of active staffs is much higher than the variation of the number of active patients. As one may guess, the main reason for this is that patients stay in the hospital all day (and night) long, while there are much less staffs in the hospital at night (only staffs of the night service S6). Moreover, this phenomenon is strengthened by a specific pattern of contacts for patients that we point below.

\begin{figure}
\centering
\includegraphics{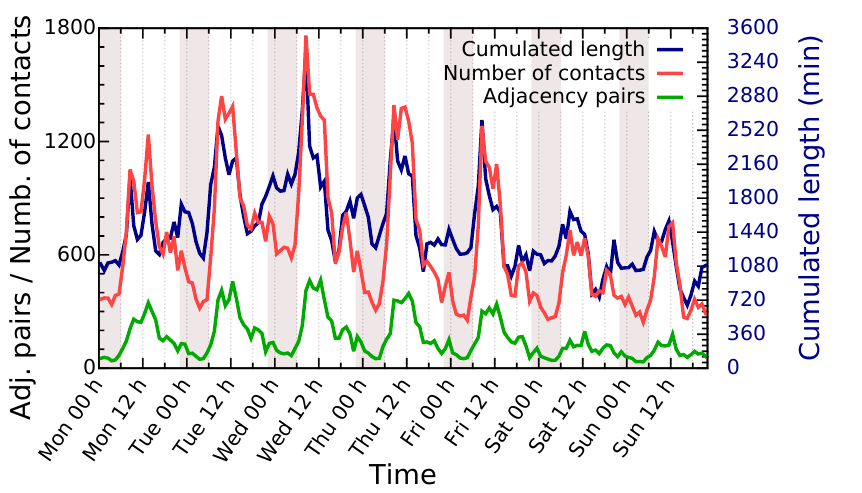}
\caption{Activity per hour in the hospital. Number of adjacency pairs (green), number of contacts (red) and cumulated length of contacts (blue).}
\label{temp-pattern-b}
\end{figure}

In Figure~\ref{temp-pattern-b}, we plot the evolution, during the same week, of the number of adjacency pairs per hour, number of contacts per hour and cumulated length of contacts per hour, in the whole hospital. All these three parameters follow the pattern highlighted above on the number of active individuals, denoting a very strong temporal structure of the activity of the hospital. Nevertheless, it is worth noticing that this pattern appears more clearly for the number of adjacency pairs and the number of contacts. The reason is that the gap between the cumulated length of contacts during daytime and during night time is not as wide as for the two other parameters. In particular, the cumulated length of contacts has high values both in the morning, between 6:00 and 12:00, and in the evening between 18:00 and 24:00. We give more explanations of this fact in the following by considering separately the activity of patients and the activity of staffs.
Let us emphasize the fact that the results we obtain here are not particular to the week we consider and hold for all the weeks of the period of study.

\begin{figure}
\centering
\includegraphics{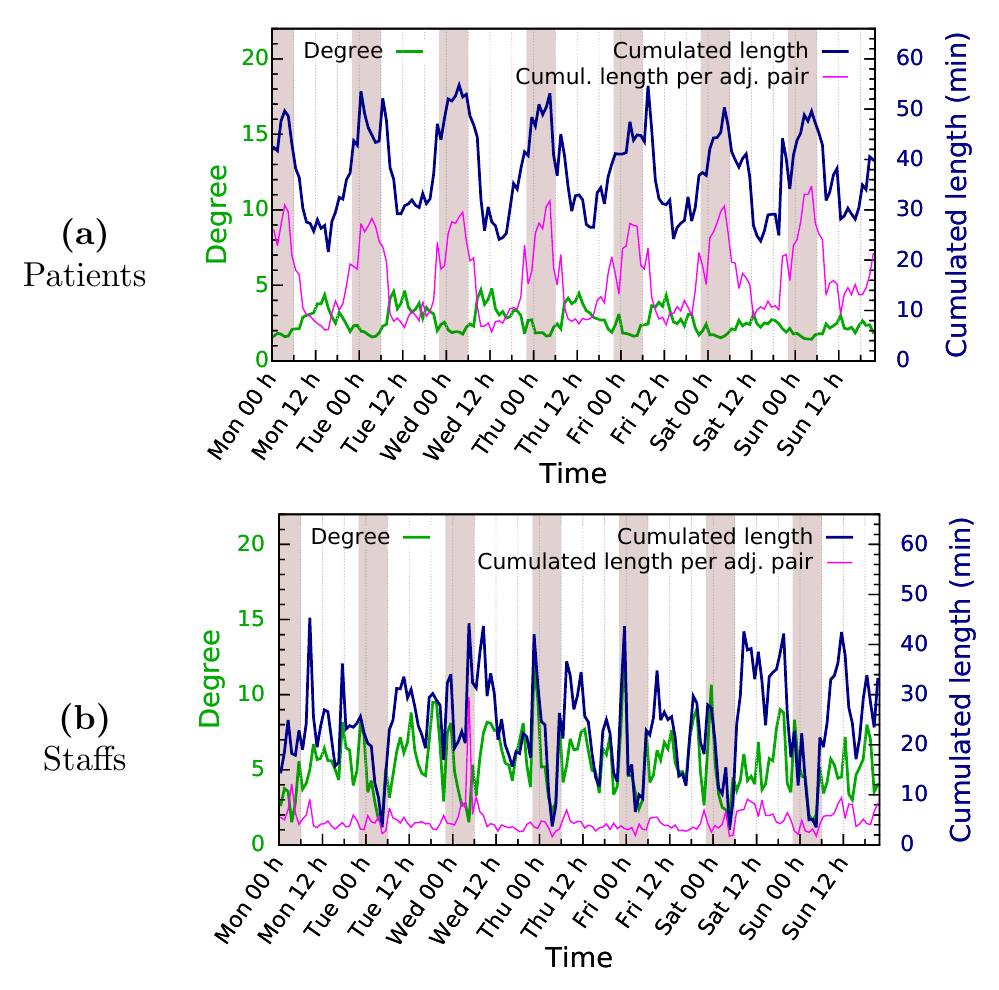}
\caption{Comparison of degree and cumulated length of contacts per hour in the hospital. (a) for active patients, (b) for active staffs. Mean degree per individual and per hour (light green), mean cumulated length of contacts per individual and per hour (dark blue) and mean cumulated length per adjacency pair and per hour (thin purple line).}
\label{temp-PA-ST}
\end{figure}

Figure~\ref{temp-PA-ST} shows the number of adjacency pairs per individual per hour (degree), cumulated length of contacts per individual per hour and the mean cumulated length of contact for an adjacency pair, separating patients (a) from staffs (b). Remember that only individuals involved in at least one contact during the considered hour are taken into account (see Definition~\ref{def:aggreg}). The reason for this is that here, we focuss on the contact pattern of the active part of the link stream.
It appears that both plots follow the circadian and weekly pattern of the whole hospital. As we pointed out previously (cf. Figure~\ref{TotalContactsPAPE}), the average cumulated length of contacts for patients is much higher than the one of staffs. On the other hand, for staffs, this cumulated length varies more in time than the one of patients: there are several peaks of activity for staffs in one day while there is mainly one for patients, moreover, the value for staffs regularly becomes very low (less than 5 minutes per hour) while the value for patients almost always remain above 10 minutes per hour. Looking closer at the times when the peaks of activity occur, one can see that, surprisingly, the cumulated length of contacts of active patients is higher during night (between 22:00 and 8:00) and much lower during days (between 10:00 and 20:00). For active staffs, the situation is opposite. Their cumulated length of contacts is very low at night (between 0:00 and 6:00) and their peaks of activity generally occur around 8:00, 12:00, 20:00 and 23:00. This observation has to be tempered with the fact that only active patients are taken into account and that their number is lower during nights, see Figure~\ref{temp-pattern-a}.

Nevertheless, interestingly, this difference can be further explained considering the curves of number of pairs and mean cumulated length per pair. For patients, the cumulated length of contacts varies opposite to the number of adjacency pairs, but follows the mean cumulated length per pair. During day time, patients are in contact with several persons and only very few at night time. But at night, the mean cumulated length of their adjacency pairs is longer. The reason for this is that most of them share their room and therefore have very long contacts with their room-mate at night, as the distance between beds is usually no more than 1.5 meters.
For staffs the situation is very different. Their mean cumulated length of contact per adjacency pair appears to be much more stable along time than the one of patients. Moreover, the variations of their cumulated length of contacts are not opposed to the variations of their number of adjacency pairs (as it is the case for patients) but are rather in accordance with them, despite the fact that they also show some visible differences.

\begin{figure}
\centering
\includegraphics{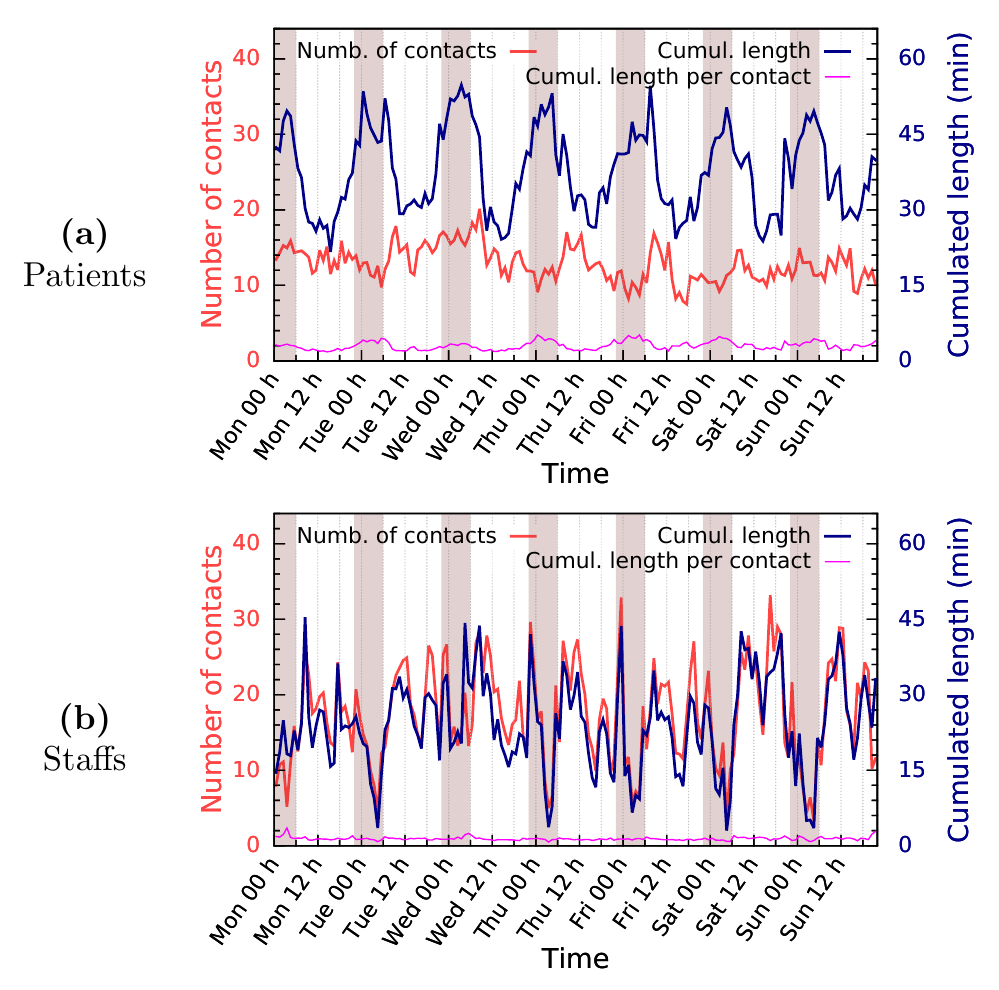}
\caption{Comparison of number of contacts and cumulated length of contacts per hour in the hospital. (a) for active patients, (b) for active staffs. Mean number of contacts per individual and per hour (light red), mean cumulated length of contacts per individual and per hour (dark blue) and mean cumulated length per contact (thin purple line).}
\label{temp-cont-PA-ST}
\end{figure}

The activity pattern of staffs appears clearly on Figure~\ref{temp-cont-PA-ST}. It shows, separately for active patients (a) and active staffs (b), the variations of the cumulated length of contacts compared with those of the number of contacts and mean duration per contact. One can notice that, opposite to the cumulated length of contacts, the number of contacts is higher for staffs than for patients. On Figure~\ref{temp-cont-PA-ST} (b), it is striking to see that the cumulated length of contacts of staffs very closely follows their number of contacts. In the meanwhile, the mean duration of their contacts appears quite stable along time with an average value lower than the one of patients (see Figure~\ref{temp-cont-PA-ST} (a)). This shows that for staffs, the cumulated length of their contacts is made by the repetition of numerous contacts (with many different persons, see Figures~\ref{temp-PA-ST} and the discussion above) and not by longer contacts with a few number of persons, as it is the case for patients. Figure~\ref{temp-cont-PA-ST} (a) confirms this fact: the cumulated length of contacts of patients does not follow at all their number of contacts, but is in accordance with the mean duration of their contacts.

Thus, the daily pattern of contacts for patients and staffs is drastically different. For patients, their cumulated length of contacts depends on the cumulated length of their adjacency pairs: it is high during nights when they have long contacts with a very restricted number of persons (usually only one or two). For staffs, their mean cumulated length of contacts per adjacency pair as well as the mean duration of their contacts do not vary much: they have longer time of contacts when they have more numerous contacts, which happens several times per day at rather fixed times around 8:00, 12:00, 20:00 and 23:00. These deep differences are very likely to have a strong impact on the way patients and staffs can propagate spreadings in the hospital.

\section*{Conclusion and perspectives}

We presented here the first analysis of the link stream of contacts in a whole hospital during a long term period. We designed a method to investigate both the temporal and topological structures of this link stream. Our method constitutes a diagnostic tool that can be used to reveal the main characteristics of the contacts with regard to the organisation of the hospital within services and socio-professional categories. It is generic and can be applied to any hospital, and more generally to any link stream where nodes are \emph{a priori} partitioned into functional groups.

The application of this method to the link stream of contacts in the hospital of Berck-sur-mer provides some important observations for understanding and controlling the spread of nosocomial infections in hospitals. First, contacts are not at all uniformly shared between services and between socio-professional categories. The analyses we conducted following these two dimensions show a very specific and strongly marked structure of the contacts in the hospital. This pattern of contacts, which may be different for each considered hospital but which also probably exists for each of them, certainly plays a key role in the propagation of infections within the hospital. Another observation, is that this pattern is different depending on whether one considers daily adjacency pairs or cumulated length of contacts. This points out that it is crucial to clarify the impact of duration of contacts on the possibility of transmission, as depending on the importance of this impact, the possibilities of diffusions within the hospital may appear quite different. Our analyses confirm that patients and staffs exhibit a quite different pattern of contacts. But contrary to what was observed in some earlier measurements~\cite{VBC+13}, in terms of cumulated length, the contacts between patients play a very important role for connecting the whole hospital, both inside services and between services.

Moreover, from a temporal point of view as well, the structure of the contacts in the hospital appears complex and clearly marked. It exhibits a strong heterogeneity which certainly has a great impact on diffusions within the hospital. Then, any policy aiming at limiting such diffusions should imperatively take into account the temporal dimension of contacts within the hospital. In addition, we note that the activity during nights should not be neglected: the major part of the duration of contacts of patients occurs during this time. The impact of this long duration proximity at night between patients sharing the same room must be further investigated for airborne diseases.

Clearly, the main perspective of our work is to determine the impact of the specific structure of contacts we highlighted on spreading processes. The challenge lies in the fact that the dataset contains only the contaminated nodes and their time of contamination, as it is impossible, with current technology, to measure through which contacts the contamination occurred. Therefore, a first step toward this goal is to use synthetic diffusion processes on the link stream of contacts to determine what are the contacts more likely to propagate diffusions, and then check whether contaminated nodes of the dataset are more often involved in such contacts than non-contaminated nodes are.

\section*{Acknowledgements}

The authors thank all the I-Bird (Individual-Based Investigation of Resistance Dissemination) study group members.


{
\bibliographystyle{plain}
\bibliography{dynamic-hospital}
}

\end{document}